**F. Huvelin, S. Dequand,**
**A. Lepage, C. Liauzun**
(ONERA)

E-mail: fabien.huvelin@onera.fr

# On the Validation and Use of High-Fidelity Numerical Simulations for Gust Response Analysis

Specific gust response is considered as one of the most important loads encountered by an aircraft. The Certification Specification (CS) 25, defined by the European Aviation Safety Agency (EASA), and the Federal Aviation Regulations (FAR) 25, defined by the Federal Aviation Administration (FAA), describe the critical gusts that an aircraft must withstand. They must be analyzed for a large range of flight points (Altitude and Equivalent Air speed) and mass configurations. For some load cases, the standard tools could not be accurate enough to correctly predict the gust response and the use of high-fidelity computation could be required. Therefore, ONERA has implemented in its in-house Computational Fluid Dynamics (CFD) code *elsA* (ONERA-Airbus-Safran property) the capability to compute the high-fidelity aeroelastic gust response, directly in the time-domain, for different discrete gust shapes.

This paper presents some recent work achieved at ONERA concerning high-fidelity simulations for gust response. First, a physical validation of the gust response simulation is performed by comparing the results to those obtained experimentally on a scaled model. Second, numerical comparisons are performed using various techniques, in order to model the gust. Finally, an application for generic regional aircraft is shown.

## Introduction

An important prerequisite for the certification of an aircraft design is to investigate the effects produced by atmospheric disturbances. In particular, the aircraft has to be designed to withstand loads resulting from gusts. One step to assess the aircraft gust response is to apply the criteria defined for certification [11], [12], [13]. One of these criteria, called "discrete gust design", considers that the airplane is subjected to symmetrical vertical and/or lateral gusts. Dynamic gust analyzes usually rely on linear techniques in the frequency domain, based on simple Doublet Lattice Methods (DLM) for the aerodynamic flow prediction [1]. These techniques are valid for subsonic flows, but could sometimes be not accurate enough to obtain realistic responses in the transonic regime, characterized by strong non-linearities, such as shocks and flow separation.

Consequently, a great effort has been made to use high-fidelity tools for gust response modelling. The most natural approach is then based on the implementation of gust models directly in the CFD code and on performing time-domain simulations [20]. However, due to the very high CPU time consumption of such an approach, alternative methods to pure unsteady CFD are necessary. A first idea consists in using CFD simulations to correct the DLM [34], [8] or to build reduced-order models (ROM) [31]. Some ROM allow the physical phenomena to be coupled by taking into account flow and flight dynamics [29], as well as structural mechanics [2], to obtain the gust response of an elastically trimmed aircraft. Another method consists in linearizing the unsteady Navier-Stokes equations with respect to the gust disturbance, which is assumed to be small. This leads to a faster resolution of the flow equations, but can provide less accurate results due to the linearization assumptions [3].

ONERA has implemented in the *elsA* software the capability to compute the high-fidelity aeroelastic response to gusts. In this paper, the so-called "Field Velocity Method" (FVM) and the corresponding linearized approach are first described. Secondly, the FVM is validated by comparison with experimental results on a scaled model. Thirdly, numerical benchmarks are performed, in order to validate both approaches. Finally, an application example for gust load alleviation is presented.

## Gust Response Modeling

The certification of a new aircraft model requires the evaluation of its response to wind gusts. The FAA (with the FAR25) and the EASA (with the CS25) have defined both discrete and continuous gust velocity profiles, which are used for the certification of the aircraft [11], [12], [13]. Both vertical and lateral gusts need to be investigated. In the present study, only discrete gusts are considered. The "one-minus cosine" gust shape is defined by:

$$\begin{cases} U & = \dfrac{U_{ds}}{2}\left(1-\cos\left(\dfrac{\pi s}{H}\right)\right) \\ U_{ds} & = U_{ref} F_g \left(\dfrac{H}{350}\right)^{\frac{1}{6}} \end{cases} \qquad (1)$$

where:

- $H$ is the gust gradient (feet), defined as the distance parallel to the flight path of the airplane for the gust to reach its peak velocity, and has to be within the 30 feet to 350 feet range;
- $s$ is the distance penetrated into the gust (feet) with the condition: $0 \le s \le 2H$;
- $U_{ds}$ is the design gust velocity in equivalent airspeed;
- $U_{ref}$ is the reference gust velocity in equivalent airspeed (feet/s);
- $F_g$ is the flight profile alleviation factor.

The specification prescribes a reference gust velocity of 56 feet/s equivalent airspeed at sea level. The required reference gust velocity is reduced linearly to 44 feet/s equivalent airspeed at 15,000 feet. It can be further reduced linearly from 44 feet/s down to 26 feet/s equivalent airspeed at 50,000 feet. The flight profile alleviation factor increases linearly from the sea level value up to a value $F_g = 1$ at the maximum operating altitude. At sea level, the flight profile alleviation factor is computed as:

$$\begin{cases} F_g & = \dfrac{1}{2}\left(F_{gz}+F_{gm}\right) \\ F_{gz} & = 1-\dfrac{Maximum\ operating\ Altitude}{25000} \\ F_{gm} & = \sqrt{R_2 \tan\left(\dfrac{\pi R_1}{4}\right)} \\ F_{gm} & = \sqrt{\dfrac{MZFW}{MTOW}\tan\left(\dfrac{\pi}{4}\dfrac{MLW}{MTOW}\right)} \end{cases} \qquad (2)$$

where:

- MZFW is the Maximum Zero-Fuel Weight;
- MTOW is the Maximum Take-off Weight;
- MLW is the Maximum Landing Weight.

## High-Fidelity Modeling

The high-fidelity simulation tool developed at ONERA for aeroelastic applications is based on the *elsA* CFD solver for the flow computation [5], [15]. Over the last decade, a general framework has been developed in the optional "***Ael***" subsystem of *elsA*, giving access in a unified formulation to several types of aeroelastic simulations, while minimizing the impact on the flow solver. The available simulations cover nonlinear and linearized harmonic forced motion, steady aeroelasticity and dynamic coupling simulations in the time-domain with various structural modelling approaches. The motivation of these developments, detailed in [10], [17], [18], is to provide a numerical tool for the prediction of various aeroelastic phenomena, such as flutter or LCO and aerodynamic phenomena involving complex nonlinear flows, such as shocks, vortex flow, and flow separation. An overview of the coupled simulation system is shown in Figure 1.

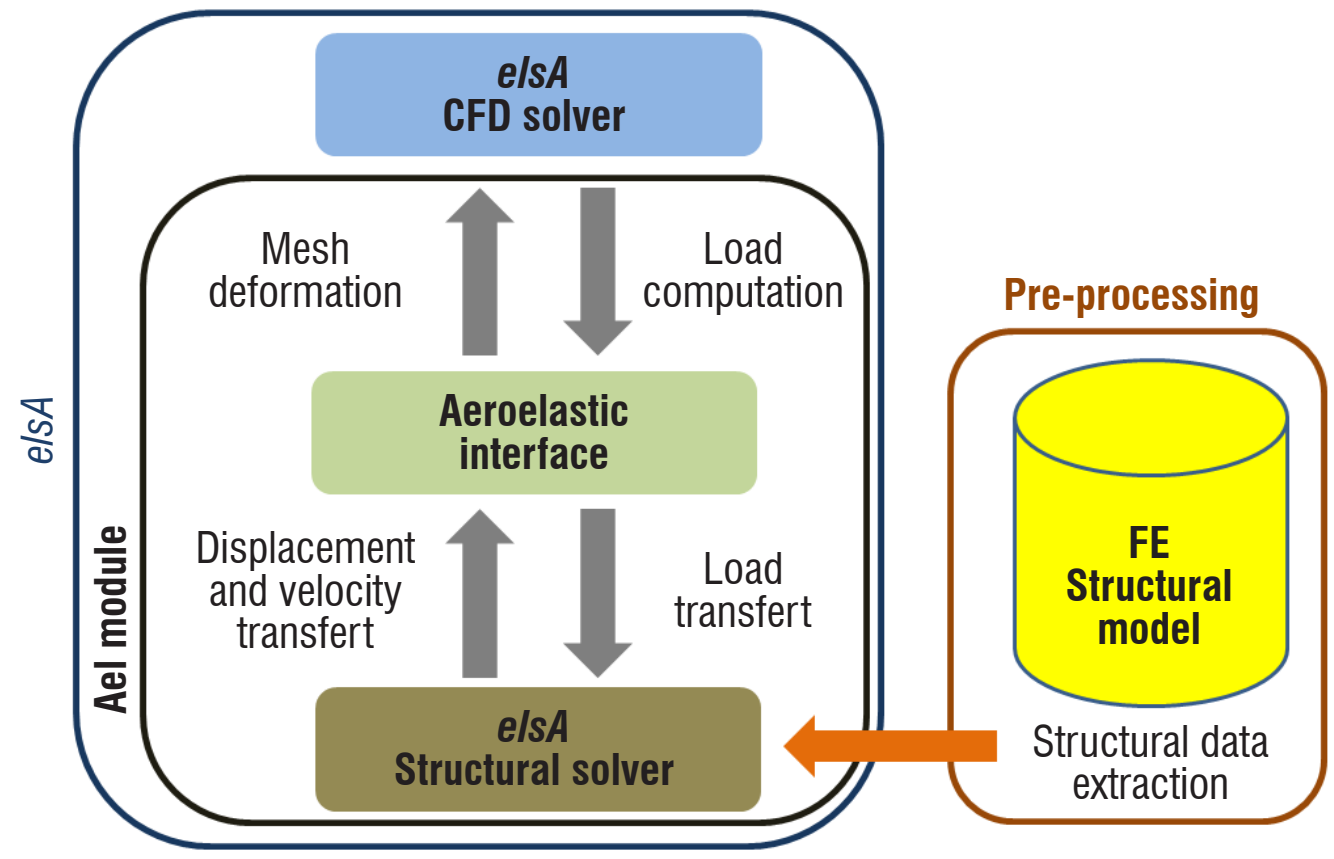

Figure 1 – Aeroelastic optional subsystem of *elsA* for aeroelastic simulations

## Field Velocity Method

There are several possibilities to implement a gust response capability in CFD codes. One way consists in introducing the gust velocity into the far field boundary conditions of the computational domain. This approach would allow not only the effect of the gust on the aircraft to be taken into account, but also the reverse effect of the aircraft on the gust [21]. However, the main drawback of such an approach is that the gust must in this case be propagated from the boundaries of the computational domain to the aircraft location, without being damped by the numerical dissipation of the discretization schemes. This would require high-order schemes and also the use of fine grids in a large part of the computational domain. An alternative to this approach is to use the so-called "Field Velocity Approach" suggested by Sitaraman *et al.* [33]. This approach takes advantage of the Arbitrary Lagrangian Euler (ALE) formulation [9], which introduces a grid velocity in the Navier-Stokes equations to take into account in a consistent way the mesh deformation in the numerical simulation.

- Mass equation

$$\left.\frac{\partial}{\partial t}\right|_{\chi} \int_{V_t} \rho dV + \int_{S_t} \rho \boldsymbol{c}\cdot\boldsymbol{n} dS = 0 \qquad (3)$$

- Momentum equation

$$\left.\frac{\partial}{\partial t}\right|_{\chi} \int_{V_t} \rho \boldsymbol{v} dV + \int_{S_t} \rho \boldsymbol{v}\boldsymbol{c}\cdot\boldsymbol{n} dS \int_{V_t} \left(\nabla\cdot\boldsymbol{\sigma} + \rho \boldsymbol{b}\right) dV \qquad (4)$$

- Energy equation

$$\left.\frac{\partial}{\partial t}\right|_{\chi} \int_{V_t} \rho E dV + \int_{S_t} \rho E \boldsymbol{c} \cdot \boldsymbol{n} dS = \int_{V_t} \left( \nabla \cdot (\boldsymbol{\sigma} \cdot \boldsymbol{v}) + \boldsymbol{v} \cdot \rho \boldsymbol{b} \right) dV \qquad (5)$$

where t is the time, $\chi$ is the reference coordinate, $V_t$ is an arbitrary volume with a surface boundary $S_t$, $\rho$ is the density, $\boldsymbol{c}$ is the convective velocity, $\boldsymbol{n}$ is the normal to the boundary surface, $\rho\boldsymbol{v}$ is the momentum, $\rho E$ is the specific total energy, $\boldsymbol{\sigma}$ is the Cauchy tensor, $\boldsymbol{v}$ is the material velocity and $\boldsymbol{b}$ is the specific body force vector. The convective velocity is expressed by the material velocity and the grid velocity ($\boldsymbol{v}_{grid}$) as follows:

$$\mathrm{c} = \boldsymbol{v} - \boldsymbol{v}_{grid} \qquad (6)$$

The standard Eulerian formulation corresponds to a grid velocity equal to 0 ($\boldsymbol{v}_{grid} = 0$), while the Lagrangian formulation corresponds to a convective velocity equal to 0 ($\boldsymbol{v} = \boldsymbol{v}_{grid}$). Due to the volume change in time, an extra conservation law has to be satisfied, the "geometric conservation law" (GCL), in order to maintain a conservative numerical scheme and to avoid additional numerical dissipation.

$$GCL: \left.\frac{\partial}{\partial t}\right|_{\chi} \int_{V_t} dV + \int_{S_t} \boldsymbol{v}_{grid} \cdot \boldsymbol{n} dS = 0 \qquad (7)$$

According to the "Field Velocity Method", a prescribed gust velocity field, depending on both space and time, is added to the grid deformation velocity in each cell of the aerodynamic grid. All equations have to be corrected with this updated grid velocity.

$$\boldsymbol{v}_{grid}(\chi,t) \rightarrow \boldsymbol{v}_{grid}(\chi,t) + \boldsymbol{v}_{gust}(\chi,t) \qquad (8)$$

The field velocity approach has been implemented in the ONERA tool *elsA-Ael* [7], [22], [26] with three discrete gust models:

- the "sharp-edged gust";
- the "one-minus cosine" profile, often used for certification;
- the "sine" profile, which could be used for the simulation of the harmonic gust response.

## Linearized Gust Response in the Frequency Domain

The high-fidelity nonlinear CFD method to compute the gust response consists in solving the URANS equations for rather long physical time durations. An alternative to this computationally expensive method is based on the linearization of the Navier-Stokes equations in the frequency domain with the fluid excited by a harmonic gust velocity. The latter derives from the linearized formulation to compute the response to a wall harmonic motion first written for turbomachinery [19] and then adapted to aircraft for load [27], [28], [30] and flutter prediction [25].

The approach implemented in the ONERA software *elsA* (LUR module, which stands for Linearized URans module) performs the linearization after having applied the space-discretization scheme. The semi-discrete URANS equations are then written using the ALE formulation to take into account the wall motion.

$$\begin{cases} \dfrac{d}{dt} \displaystyle\int_{\Omega(t)} \boldsymbol{W} d\Omega & = -\displaystyle\int_{S_t} \left( \begin{pmatrix} f(\boldsymbol{W}) \\ g(\boldsymbol{W}) \\ h(\boldsymbol{W}) \end{pmatrix} - \boldsymbol{W}\boldsymbol{v}_{grid} \right) \cdot \boldsymbol{n} dS \\ & \quad + \displaystyle\int_{\Omega(t)} \boldsymbol{T}(\boldsymbol{W}) d\Omega \\ \dfrac{d}{dt} \displaystyle\int_{\Omega(t)} d\Omega & = \displaystyle\int_{S_t} \boldsymbol{v}_{grid} \cdot \boldsymbol{n} dS \end{cases} \qquad (9)$$

where $\boldsymbol{W}$ represents the flow conservative variables ($\rho$, $\boldsymbol{v}$, $E$), $\Omega$ is the volume of the cell, $\boldsymbol{f}$, $\boldsymbol{g}$ and $\boldsymbol{h}$ are the convective and diffusive fluxes in the three space directions, $\boldsymbol{n}$ is the vector normal to the wall, $\boldsymbol{T}$ is the source term (null for the case of gust response), and $\boldsymbol{v}_{grid}$ is the grid deformation velocity vector. In the case of gust response simulation, no wall motion is considered. However, according to the Sitaraman approach [33], the gust velocity is introduced into the grid deformation velocity vector:

$$\boldsymbol{v}_{grid} \rightarrow \boldsymbol{v}_{grid} + \boldsymbol{V}_{gust}$$

The fluid variables are thereafter written as the sum of a steady or time constant part denoted by the subscript s and a perturbation part that is assumed to be harmonic of small complex amplitude:

$$\begin{cases} \boldsymbol{W} & = \boldsymbol{W}_s + \delta\boldsymbol{W} e^{i\omega t} \\ \boldsymbol{f} & = \boldsymbol{f}_s + \delta\boldsymbol{f} e^{i\omega t} \\ \boldsymbol{g} & = \boldsymbol{g}_s + \delta\boldsymbol{g} e^{i\omega t} \\ \boldsymbol{h} & = \boldsymbol{h}_s + \delta\boldsymbol{h} e^{i\omega t} \\ \boldsymbol{V}_{gust} & = e^{-i\boldsymbol{k}(\boldsymbol{X}-\boldsymbol{X}_0)} e^{i\omega t} \end{cases} \qquad (10)$$

where $\boldsymbol{k}$ is the wave number vector $\boldsymbol{k} = \frac{2\pi}{\lambda} \boldsymbol{u}_{Gust}$, $\boldsymbol{u}_{Gust}$ is the gust propagation unit vector, $\boldsymbol{X}$ represents the space coordinates of a point, $\lambda = \frac{2\pi U_\infty}{\omega}$ is the wavelength, $U_\infty$ is the aircraft flight speed and $\omega$ is the gust angular frequency. Linearizing (10) using (11) yields the complex linear system in $\delta\boldsymbol{W}$:

$$\begin{cases} i\omega\Omega_s\delta\boldsymbol{W} + \displaystyle\sum_l \frac{1}{2} \begin{pmatrix} \delta f + \delta f_l \\ \delta g + \delta g_l \\ \delta h + \delta h_l \end{pmatrix} \cdot \boldsymbol{n} \\ + \displaystyle\sum_l \boldsymbol{V}_{gust} \cdot \boldsymbol{n}_{sl} \frac{1}{2} (\boldsymbol{W}_s + \boldsymbol{W}_{sl}) = 0 \end{cases} \qquad (11)$$

The metrics (volumes and normal vectors) remain indeed constant for the case of a gust response and the fluxes perturbations $\delta\boldsymbol{f}$, $\delta\boldsymbol{g}$, $\delta\boldsymbol{h}$ are linear in $\delta\boldsymbol{W}$. The linear system is solved using a pseudo time approach based on a backward Euler algorithm with an LU-SSOR implicit stage. All acceleration techniques usually used to obtain a steady CFD RANS solution as multi-grid or local time steps can be used.

## Experimental Validation

### Experimental Set-Up

In order to generate an experimental database for the validation of high-fidelity numerical codes, a test campaign was performed in the ONERA S3Ch facility (Figure 2). This closed return wind tunnel (WT) is a transonic continuous run facility with a 0.8 m x 0.8 m square test section operating at atmospheric stagnation pressure and stagnation temperature, and is equipped with deformable adaptive walls (top and bottom walls).

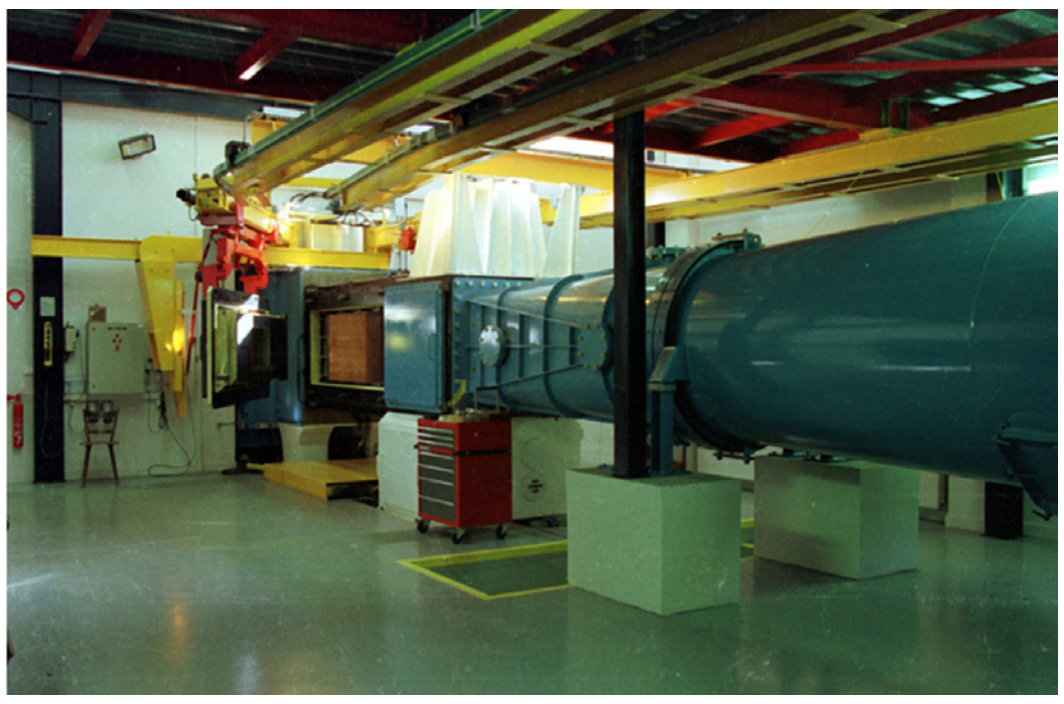

Figure 2 – The ONERA S3Ch transonic Wind Tunnel

The experimental set-up is composed of a gust generator and an aeroelastic model (Figure 3 and Figure 4). The purpose of the gust generator is to have an experimental tool able to generate relevant perturbations (gust load) for wind tunnel conditions, from the subsonic to the transonic range. The concept of the gust generator consists of two identical oscillating airfoils installed upstream of the wind tunnel test section and producing air flow deflections to generate a cylindrical gust field downstream. Its functioning is based on synchronous dynamic motions of the 2 airfoils (pitch motions) performed by 4 servo-hydraulic jacks with a frequency bandwidth of 100 Hz.

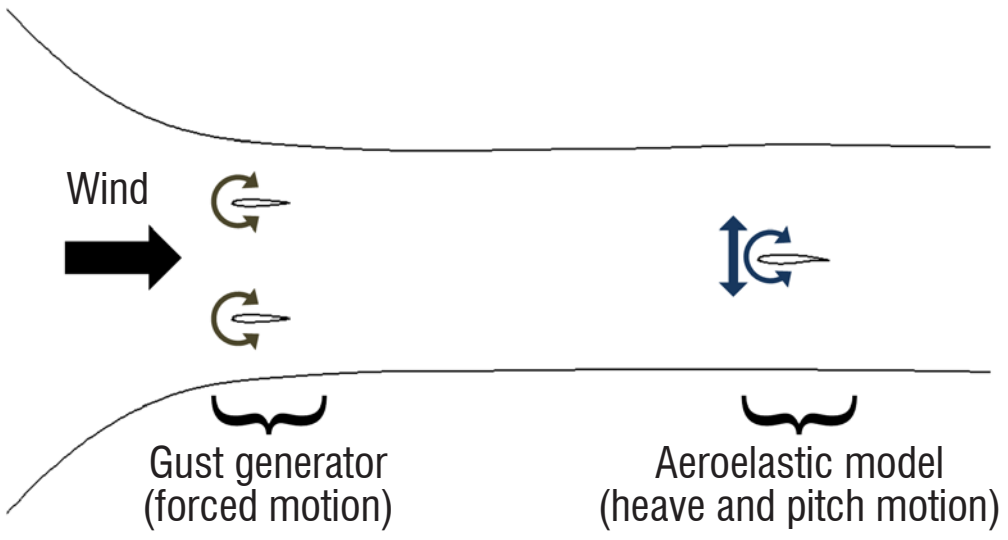

Figure 3 – Sketch of the experimental set-up

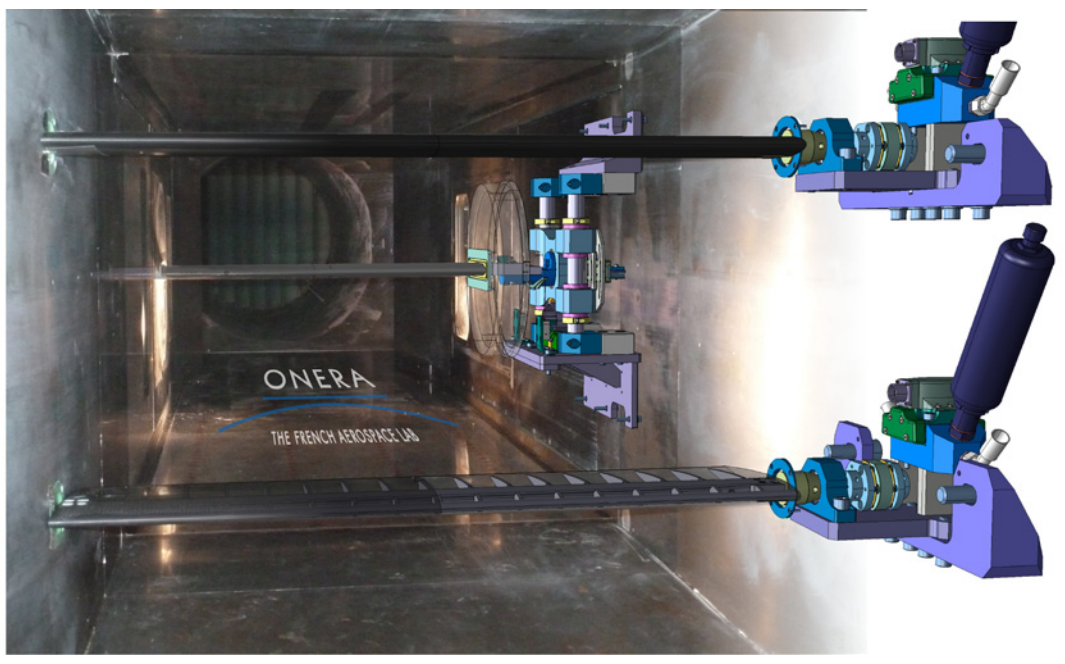

Figure 4 – "Inside Artist view" of the experimental set-up in the S3Ch facility: the gust generator (foreground) and the aeroelastic model (background)

The aeroelastic model is aimed at representing the behavior of a classical aeroelastic model with 2 degrees of freedom (dof), *i.e.*, a 2D model with heave and pitch motions. The aerodynamic part is based on the OAT15A airfoil (ONERA supercritical airfoil [32]) with a 0.25 m chord length. In order to preserve the 2D characteristic of the flow, the airfoil is designed as rigid as possible and is composed of a steel spar and 2 upper and lower carbon reinforced skins. A specific manufacturing process was defined to avoid any geometrical variations and to respect the aerodynamic shape of the airfoil (no cover, no access). The pitch and heave dof are driven by a couple of stiffness (flexible beams) and mass parameters, in addition to an arrangement of bearings in order to better "constrain/prescribe" the "rigid body" motions of the wing. In the WT test section, the mounting system is composed of 2 identical mounting parts located on each test section door. The model is equipped with a full span trailing edge control surface driven on either side by a high torque – high speed actuator allowing dynamic deflections up to 100 Hz. The instrumentation of the model is made of steady and unsteady pressure transducers, accelerometers and strain gages.

The experimental roadmap was split into several phases to correctly investigate gust load in a WT environment [24]. A first WT test campaign has been carried out to qualify the unsteady flow induced by the gust generator and its ability to generate a cylindrical gust field with significant and reproducible amplitudes in subsonic and transonic ranges [4]. Then, a second WT tests was devoted to the analysis of the gust effects on the test model behavior, *i.e.*, the aerodynamic and aeroelastic responses to an impacting gust. The final WT test objective was the demonstration in real time of gust load alleviation through the active control of the model aeroelastic response for a gust disturbance [23].

The achieved WT tests have provided a comprehensive and consistent database for the validation process of gust simulation capacities with the CFD/CSM HiFi tools.

### Physical validation

The numerical simulation allows the Field Velocity Method implemented in the non-linear equation solver to be validated.

The OAT15A airfoil was modeled with far-field conditions and conditions of adherent wall on the airfoil. A C-mesh was built around the airfoil.

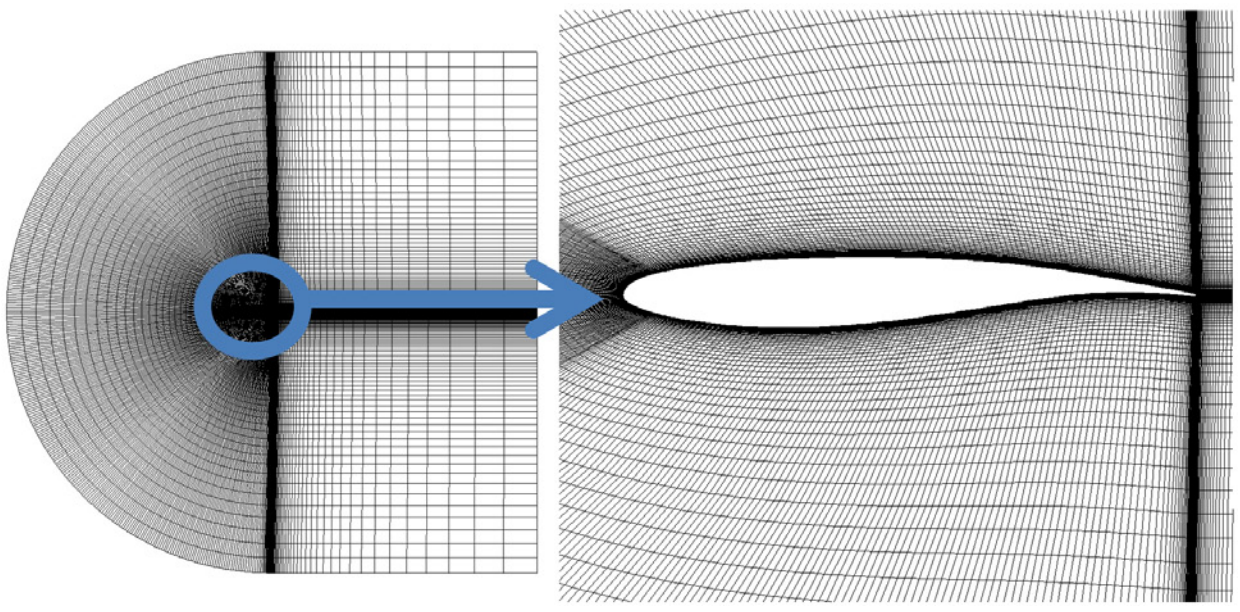

Figure 5 – Far-field mesh: overview and, OAT15 airfoil

Tests were performed at Mach number 0.73 with an angle of attack (AoA) of 2°. Numerical simulations were carried out by solving the non-linear Unsteady Reynolds-Averaged Navier-Stokes (URANS) equations. Two kinds of gust response simulations were performed. For the first one, the physical validation was performed with a rigid airfoil (fully clamped model), in order to validate the flow around the profile. The gust frequency was set to 20 Hz. For the second one, the airfoil was able to move according to its heave and pitch degrees of freedom, thus allowing the assessment of its aeroelastic response and its comparison with experimental results. The gust frequency was set to 25 Hz (frequency of the heave mode). For both simulations, a steady state was first computed. The aerodynamic parameters were adjusted to fit the steady experimental results.

The numerical and experimental unsteady pressure distributions were compared using 38 unsteady pressure transducers located along the center line in the spanwise direction. The same Fourier analysis was applied to both the numerical and the experimental results, in order to avoid any additional or compensation errors.

For the rigid response, a good agreement between the numerical and experimental data can be observed for the magnitude of the pressure (Figure 6). The extrema are indeed correctly predicted. With regard to the phase, a good agreement is also noticed on the upper surface up to the shock. However, larger differences arise close to the trailing edge.

For the aeroelastic response, the aerodynamic flow around the airfoil is well predicted (Figure 7). The pressure magnitude peak is appropriately captured, thanks to the tuning of the FVM model. The latter model indeed avoids numerical dissipation and cannot take into account the physical dissipation of the flow perturbation generated by the gust generator. The numerical gust amplitude encountered by the airfoil must then be determined according to the flow velocity measurements provided by the probe located ahead of the leading edge. The unsteady pressure magnitude after the shock root and on the lower surface is less accurately predicted. A large difference appears in the phase around the trailing edge.

The numerical and experimental acceleration were compared using 4 accelerometers located along the chord.

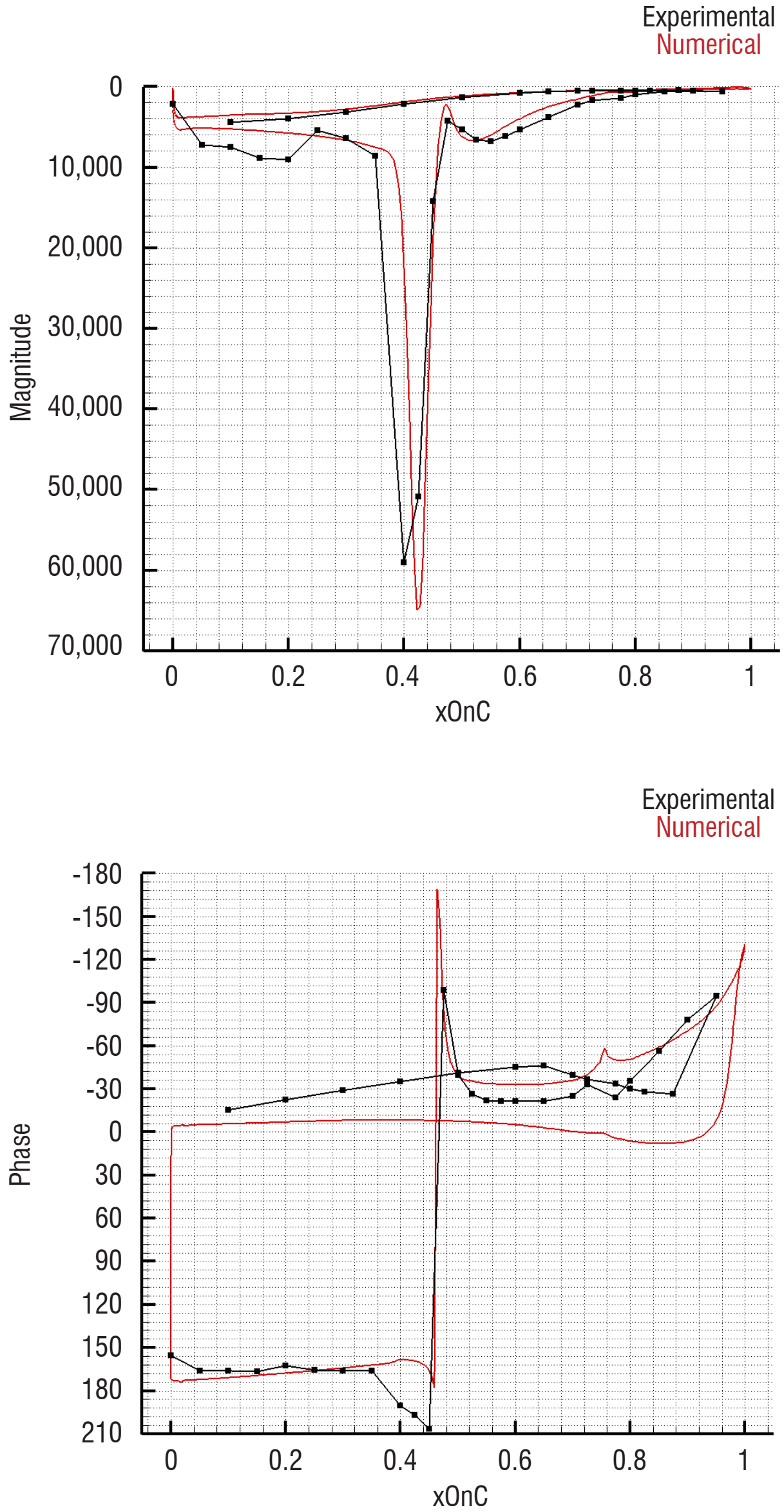

Figure 6 – Comparison between numerical and experimental unsteady pressure distributions – rigid airfoil

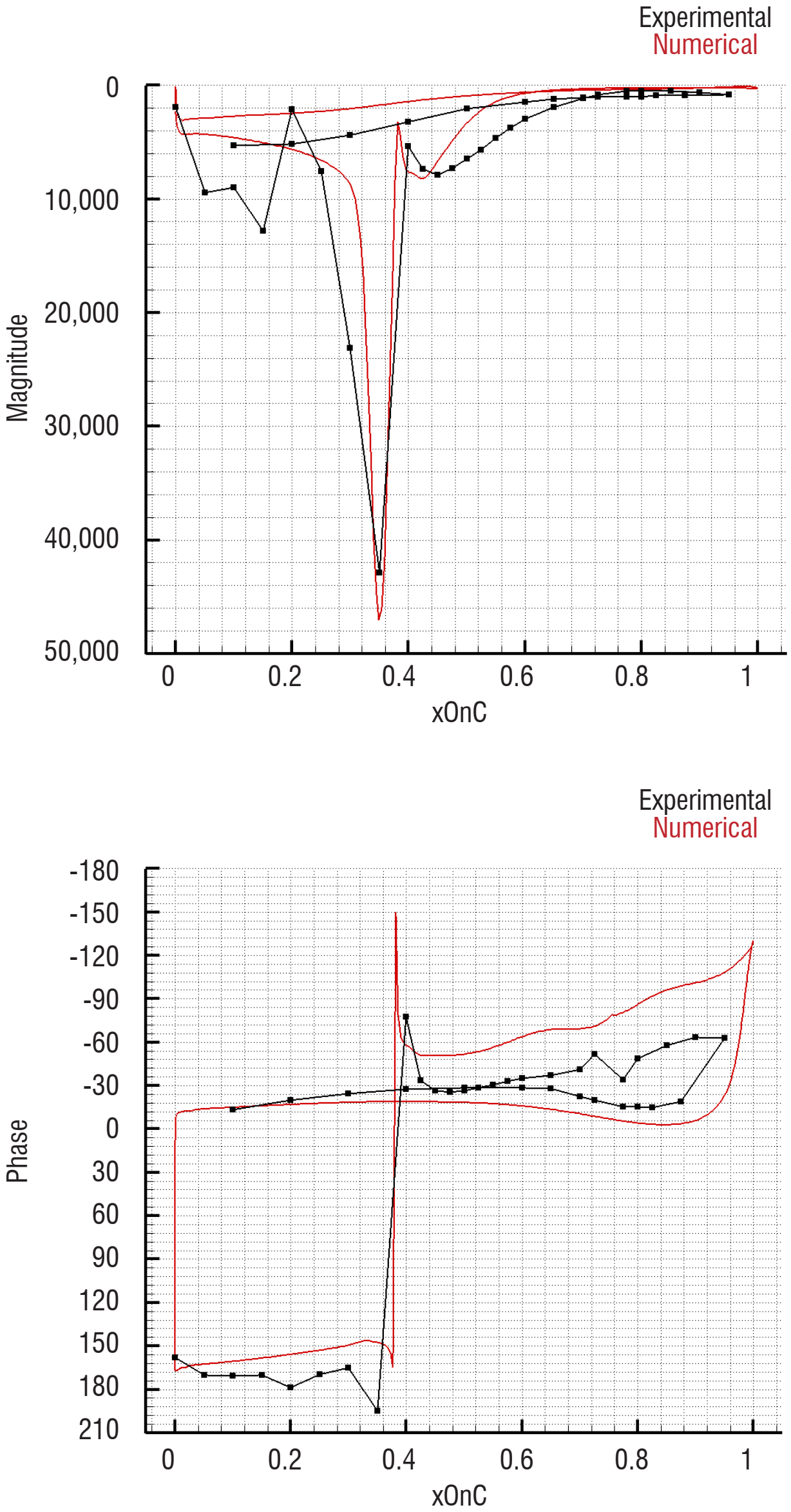

Figure 7 – Comparison between numerical and experimental unsteady pressure distributions – aeroelastic airfoil

The structural behavior is better predicted (Figure 8) than the pressure distribution. The magnitude of the acceleration is closer to the experimental result near the leading edge. A difference appears around the third probe due to the hinge of the control surface present in the mock-up. The hinge is not stiff and its flexibility is not taken into account in the computation. The phase is accurately predicted.

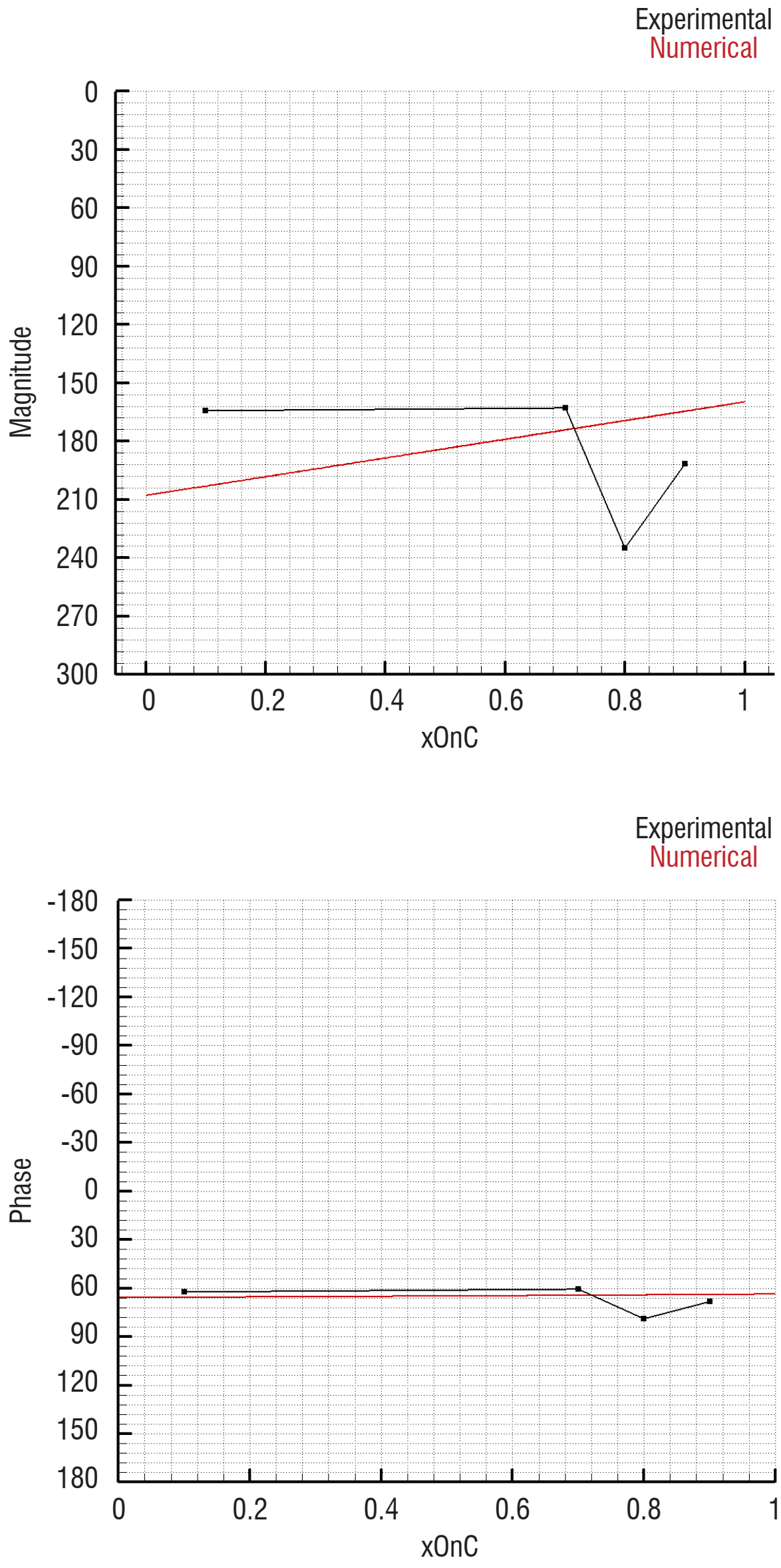

Figure 8 – Comparison of structural responses between numerical and experimental results – aeroelastic airfoil

## Numerical validation

### One-scale numerical benchmark for FVM

Dynamic gust analyzes usually rely on linear techniques in the frequency domain, based on simple Doublet Lattice Methods (DLM) for the aerodynamic flow prediction. These techniques are valid for subsonic flows, but could sometimes be not accurate enough to obtain realistic responses in the transonic regime, characterized by strong non-linearities such as shocks and flow separation. This method is compared to the high-fidelity approach using the Field Velocity Method on an industrial case. The Airbus XRF-1 transport aircraft configuration has been used for the benchmark. It is a generic research configuration representative of wide-body modern civil transport aircrafts.

The structured aerodynamic mesh was built around a cruise shape and includes about 7.47 million cells (Figure 9). The grid designed for URANS simulations, is thus rather coarse for a half-aircraft configuration. It is therefore not possible to accurately capture the viscous phenomena, especially around the nacelle, on which a wall slip condition was therefore applied. An adiabatic condition of adherent wall was applied everywhere else on the aircraft. All of the RANS and URANS simulations were performed using the Spalart-Allmaras turbulence modeling.

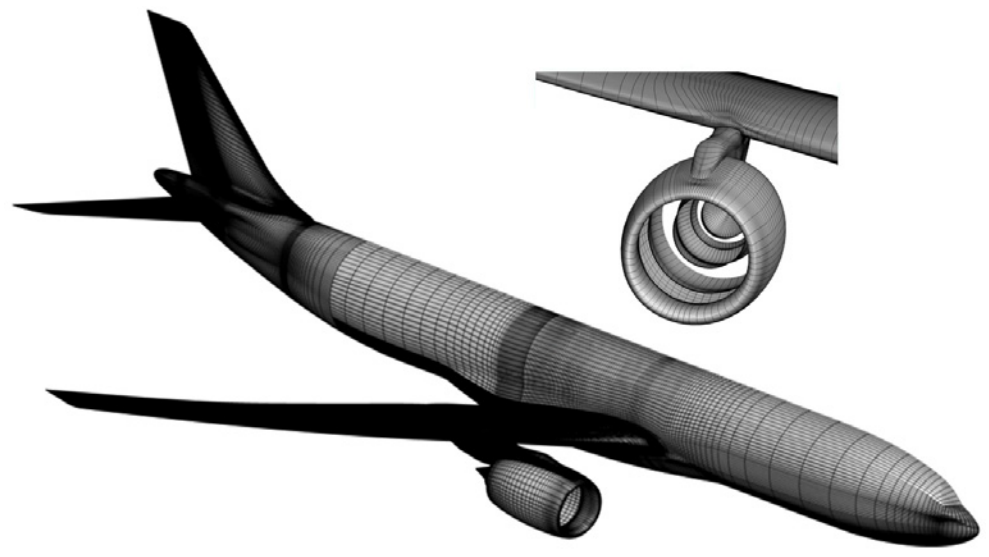

Figure 9 – Aerodynamic mesh

A simplified Nastran Finite-Element model of the whole aircraft was built, based on a detailed representation (solid, shell and bar elements) of the wings and the central part of the fuselage, and on condensed elements (super elements) for the front and rear parts of the fuselage and for the tail (Figure 10). For CFD and DLM aeroelastic simulations, a structural damping ratio equal to 2% of the critical damping was imposed.

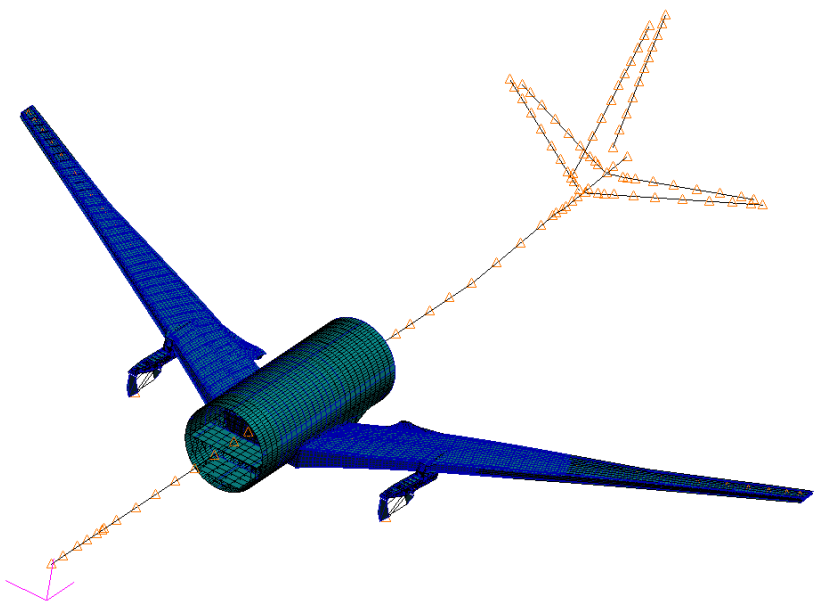

Figure 10 – Structural model

#### Case and benchmark description

The capability to predict the response to a gust has been assessed for typical cruise flight conditions (Mach = 0.86, Altitude = 35,000 ft, Mass = 230 tons and AoA = 2.3°). The applied discrete gust velocity corresponds to a "one-minus cosine" shape and the parameters of the gust were selected using the FAR25 rules. In particular, the gust gradient ($H$ = 350 ft), and the design gust velocity ($U_{ds}$ =9.82 m/s) were evaluated from the mass and altitude features of the aircraft. The gust induced angle of attack corresponds to $\Delta\alpha$ = 2.28° at the peak. A physical time duration of 4.0 s was simulated.

Several simulations were performed in order to study dynamic gust responses. Both rigid and flexible high-fidelity gust dynamic simulations were run with *elsA*, in order to quantify the effect of flexibility on the load distribution. A Nastran gust response simulation was also run using, as the high-fidelity approach, a restrained aircraft hypothesis. The objective of this computation is first to validate the high-fidelity approach, and also to investigate its benefits with respect to the linear aerodynamic Doublet Lattice Method approach used in Nastran.

**Result comparison**

Figure 11 shows the additional load factor ($\Delta N$) for three computations, *i.e.*, CFD rigid (rigid load factor), CFD aeroelastic (aeroelastic load factor) and Nastran DLM (Nastran load factor), corresponding to the Maximum Take-Off Weight (MTOW) load case. The gust amplitude time history is also plotted.

*elsA-Ael* and Nastran simulations predict a similar maximum load factor, with a phase shift with respect to the gust input. After the first cycle, the two dynamic responses differ, with larger unsteady levels in the non-linear *elsA-Ael* simulation but a similar pseudo-frequency.

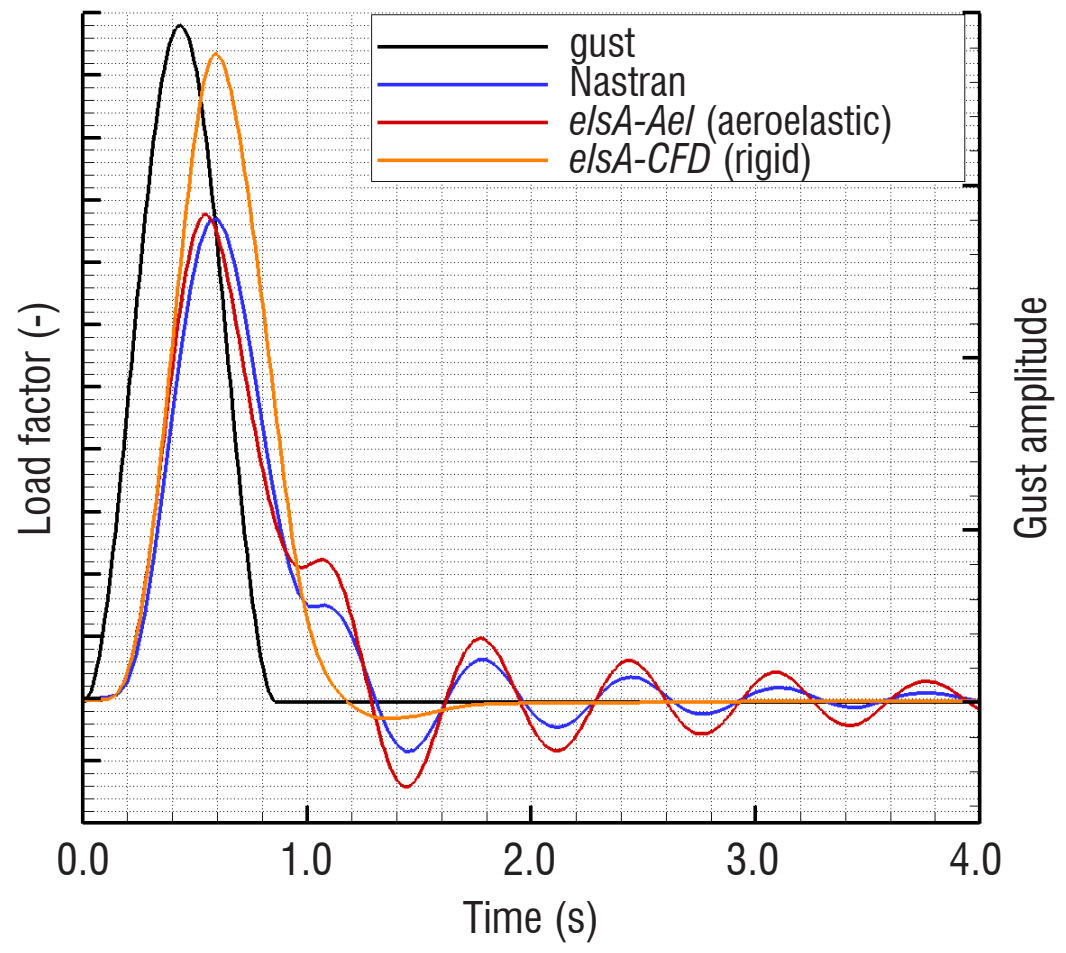

Figure 11 – Time evolution of additional gust load factor

The rigid computation predicts a higher maximum load factor with a greater delay than that of the aeroelastic simulations. The reason for this over-estimation is mainly due to the inertial forces, which are only taken into account in the aeroelastic simulations. Indeed, the inertia relief has a favorable effect on the load factor. To check this assumption, an aeroelastic Nastran computation for an OWE configuration has been achieved. OWE results lead to a higher maximum induced load factor than that of the rigid MTOW, showing the beneficial effect of inertia relief for the MTOW case. Figure 12 shows the maximum displacement and twist over time for the nonlinear *elsA-Ael* and Nastran MTOW computations.

The pseudo-frequencies predicted by both computations are roughly identical. Larger displacements are, however, observed in the case of the nonlinear aerodynamics of the *elsA-Ael* computation, leading to a difference of damping between the two computations.

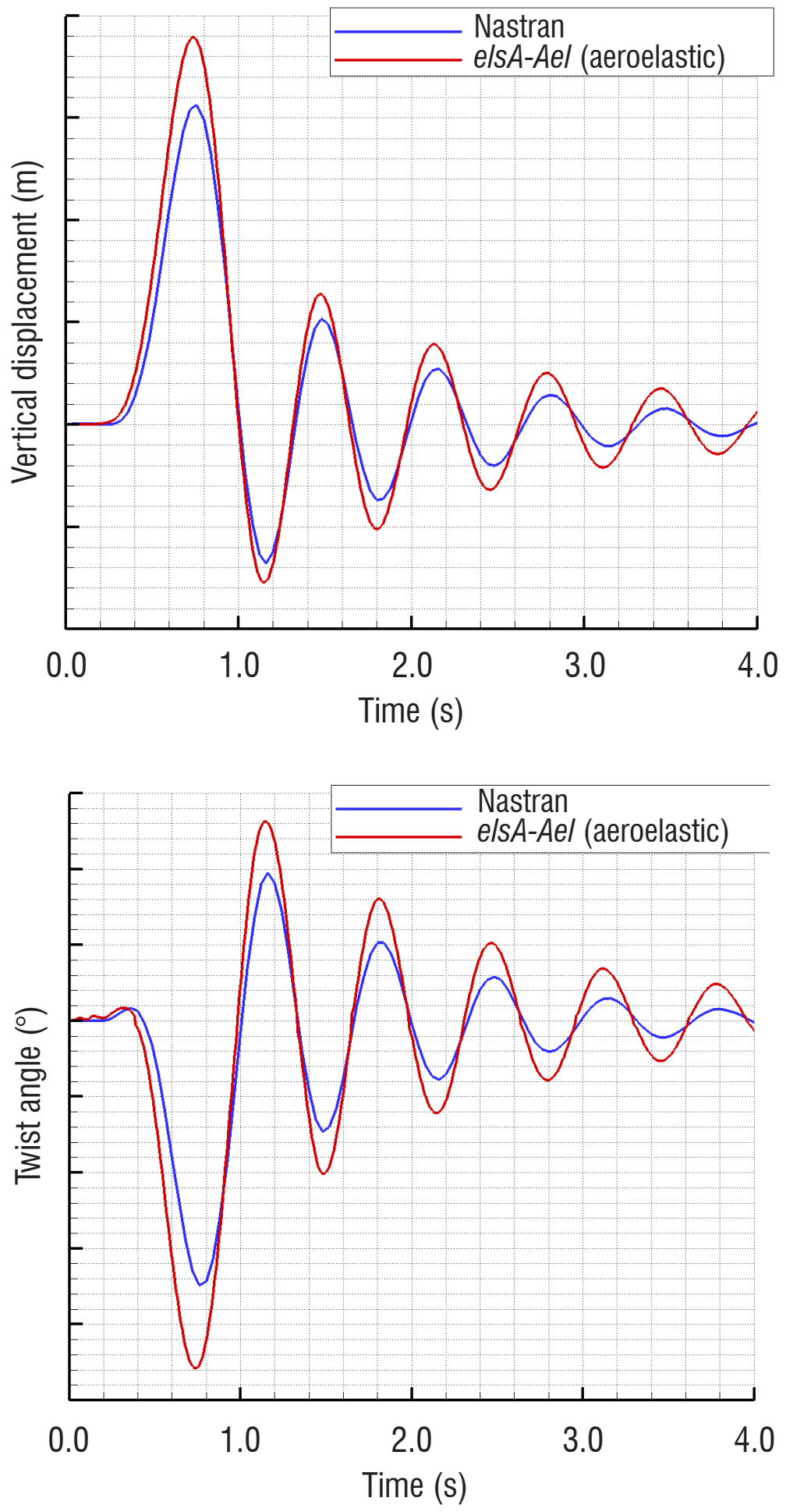

Figure 12 – Time-evolution of maximum vertical displacement (up) and twist (bottom)

The combined load diagram (Figure 13) shows that the maximum/minimum of the twisting moment corresponds to the maximum/minimum of the shear force for both rigid and aeroelastic computations. The rigid simulation exhibits a higher maximum shear force and twisting moment than the aeroelastic simulation. *elsA-Ael* and Nastran estimate a similar minimum and maximum transverse force, while the twisting moment is under-estimated by the Nastran calculation in comparison with *elsA-Ael*.

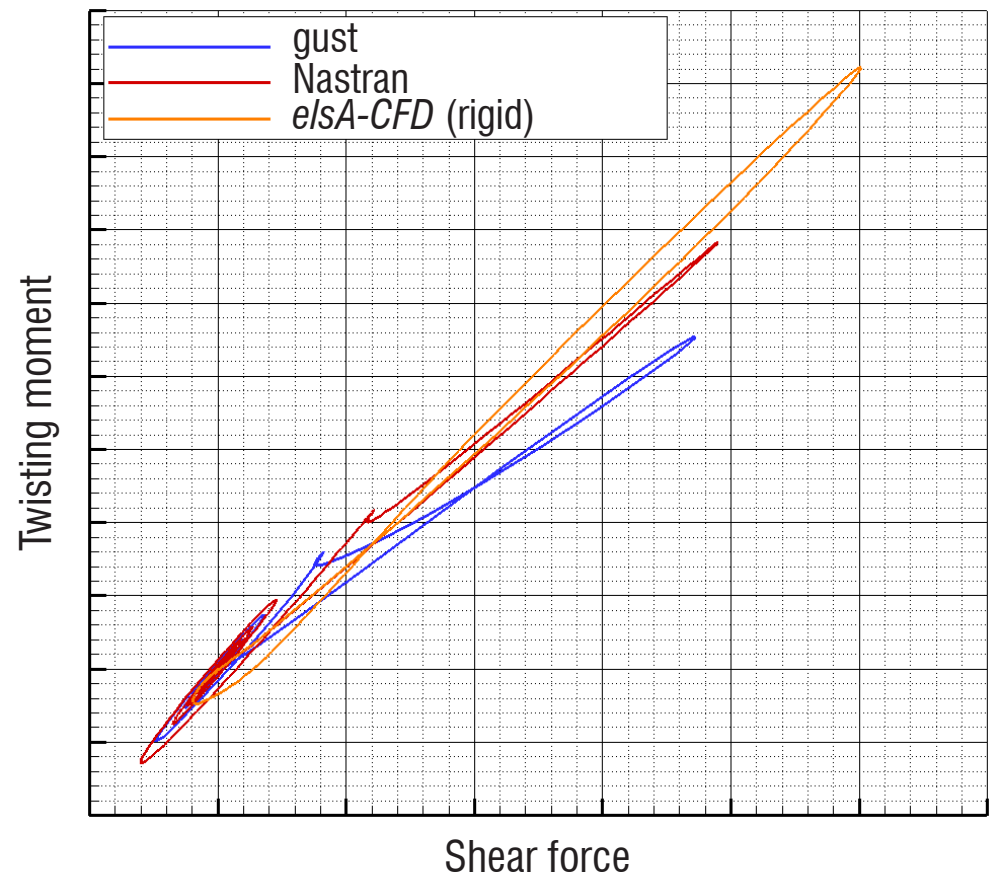

Figure 13 – Time-evolution of combined loads plot at the wing root for rigid, aeroelastic and NASTRAN computations

The gust has a large impact on the flow distribution around the wing (Figure 14). Before the gust encounter, the flow over the wing is rather two-dimensional. When the gust reaches the wing, a disturbance appears at the tip part and expands towards the wing root until a flow separation occurs. After the gust encounter, the flow returns to its initial state.

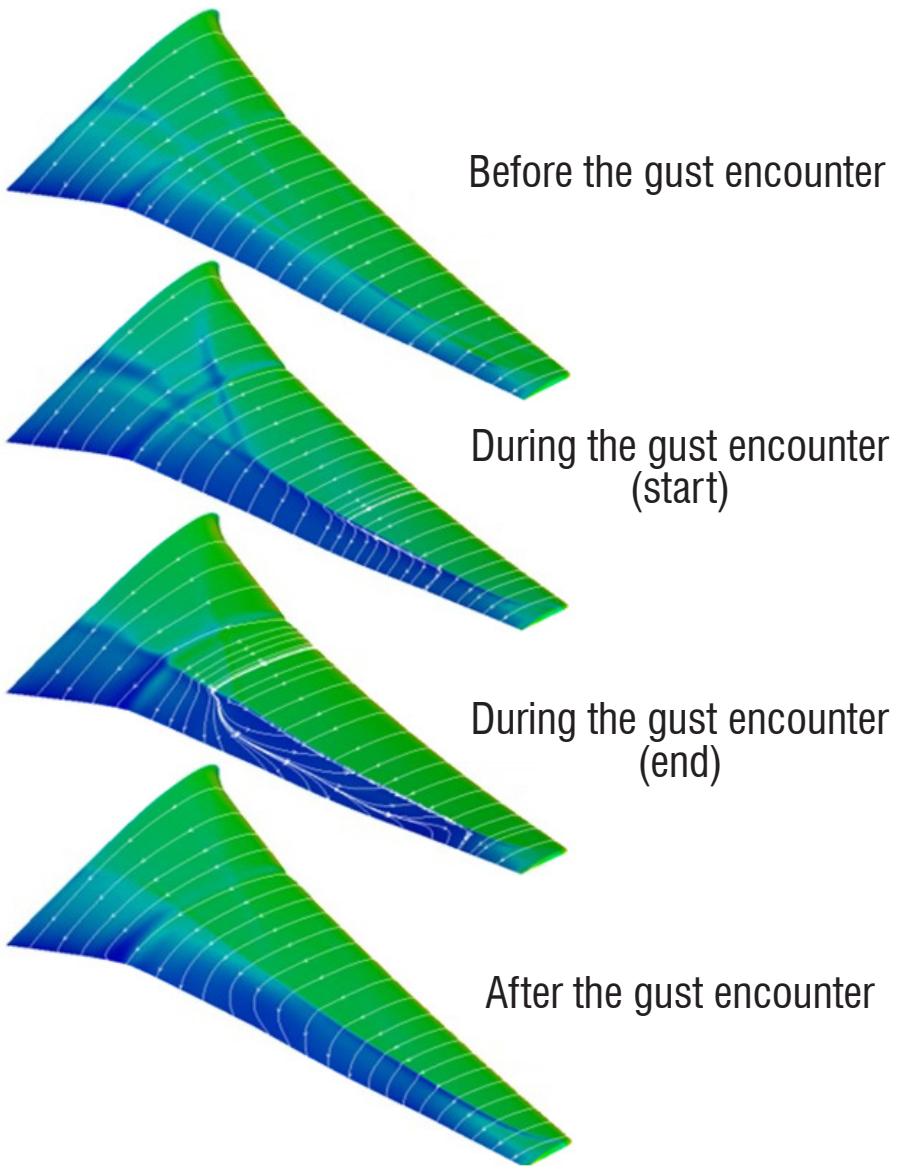

Figure 14 – Friction stream traces for the *elsA-Ael* simulation before, during and after the gust encounters the wing

## Numerical 2D benchmark for the linearized approach

The linearized approach has been validated in an industrial context to compute gust loads in the subsonic regime [35]. This benchmarks is aimed at validating the approach for transonic flight conditions. Given that the Field velocity method has been validated by comparing experimental and numerical results, this method has been used as reference, in order to validate the linearized formulation.

This alternative method to compute the gust response has been assessed by comparisons with non-linear URANS simulations for two cases: the 2D airfoil NACA64A010 in a transonic viscous flow and a 3D wing in an inviscid flow.

Gust responses of the 2D symmetric airfoil have been computed for transonic conditions for which experiments have been carried out for both steady and harmonic pitching motion measurements [6].

$$\begin{cases} Mach & = & 0.796 \\ P_i & = & 203321\,Pa \\ T_i & = & 310\,K \\ \alpha & = & 0^o \end{cases}$$

As a first step, a steady simulation was performed using the Spalart-Allmaras turbulence model, and yielded a well-converging solution exhibiting, as expected, a strong shock (Figure 15).

Responses to harmonic gust excitations of a wavelength 25 times the chord matching a gust frequency of 21.17 Hz were computed using both the linearized and non-linear URANS solvers. The non-linear simulations were carried out for gust amplitudes $V_G = \frac{U_\infty}{60}$ matching an incidence variation of 0.955°, $V_G = \frac{U_\infty}{300}$ (0.191°) and $V_G = \frac{U_\infty}{1500}$ (0.0382°). They were run for physical time durations long enough to reach the harmonic regime, as can be seen in Figure 16.

Unlike the simulation with the lowest gust amplitude (25 times lower), the one with the largest amplitude exhibits a large shock motion inducing probably unsteady non-linear phenomena in the flow field (Video 1 and Video 2). Since the linearized solver actually provides the sensitivity of the unsteady pressure to the input excitation (here the gust amplitude), comparisons with the nonlinear solvers have been carried out on the first harmonic of the Fourier Series of the pressure coefficient time signal divided by the gust amplitude. Figure 17 shows these complex unsteady Cp distributions resulting from both linearized and URANS simulations. The distributions obtained with the non-linear solver indeed tends with decreasing gust amplitude to the distribution obtained using the linearized solver, which validates the linearized formulation for the 2D symmetric airfoils in transonic flows and confirms the nonlinear unsteady phenomena occurring with the largest gust amplitudes.

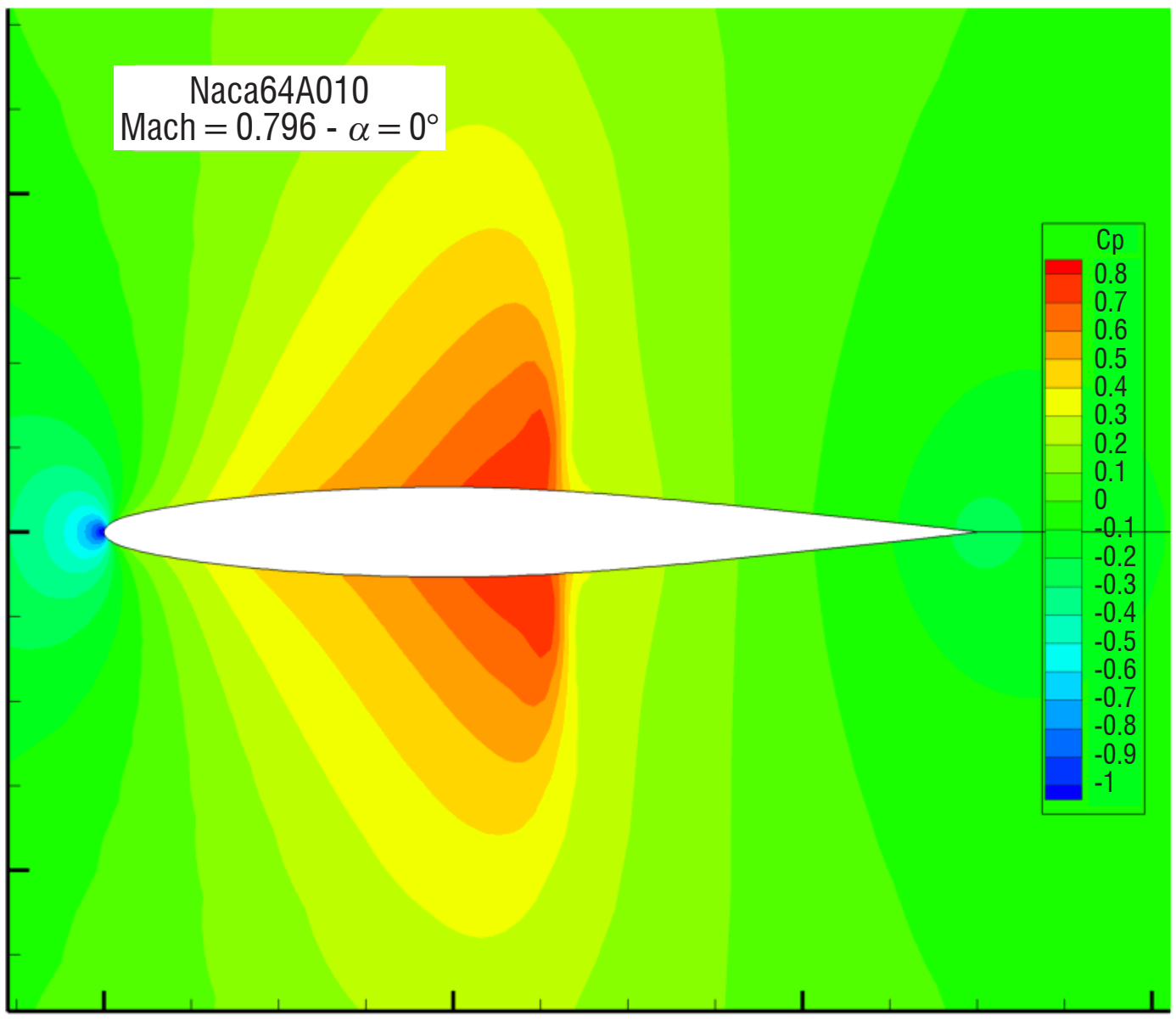

Figure 15 – Steady Cp distribution

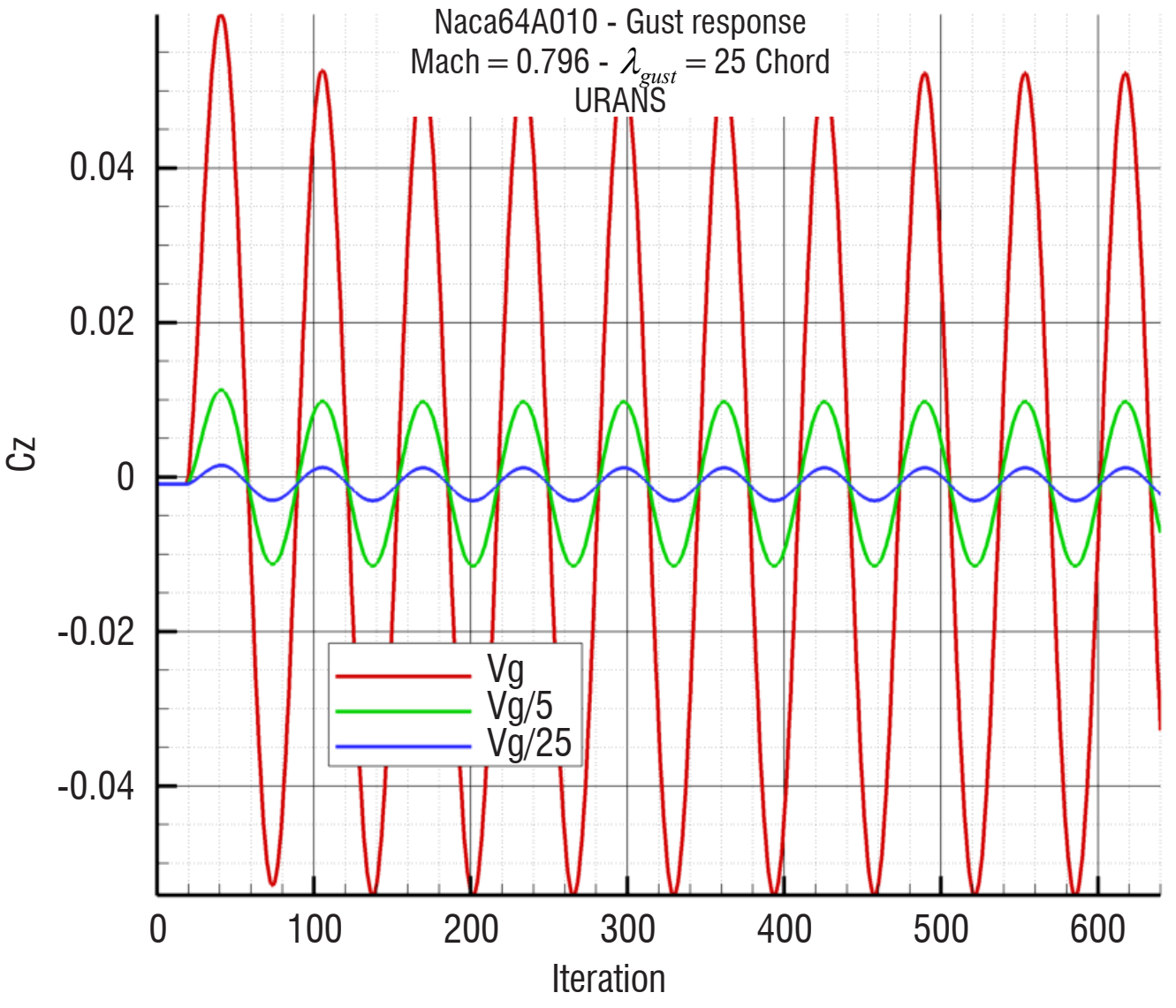

Figure 16 – Histories of Lift coefficient computed with the non-linear solver

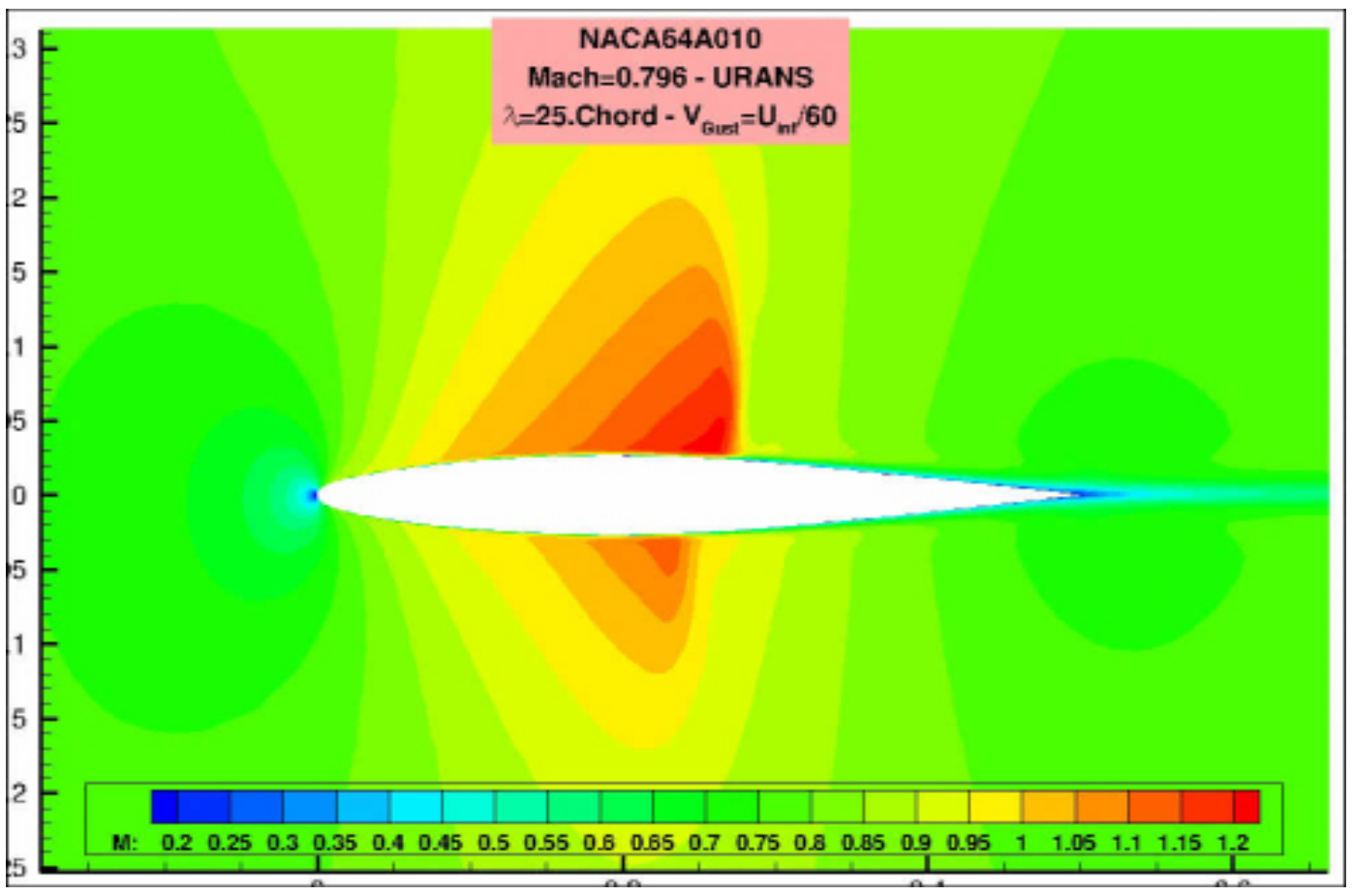

Video 1 – Mach distributions for a gust period (URANS simulation with the gust amplitude equal to U∞/60)

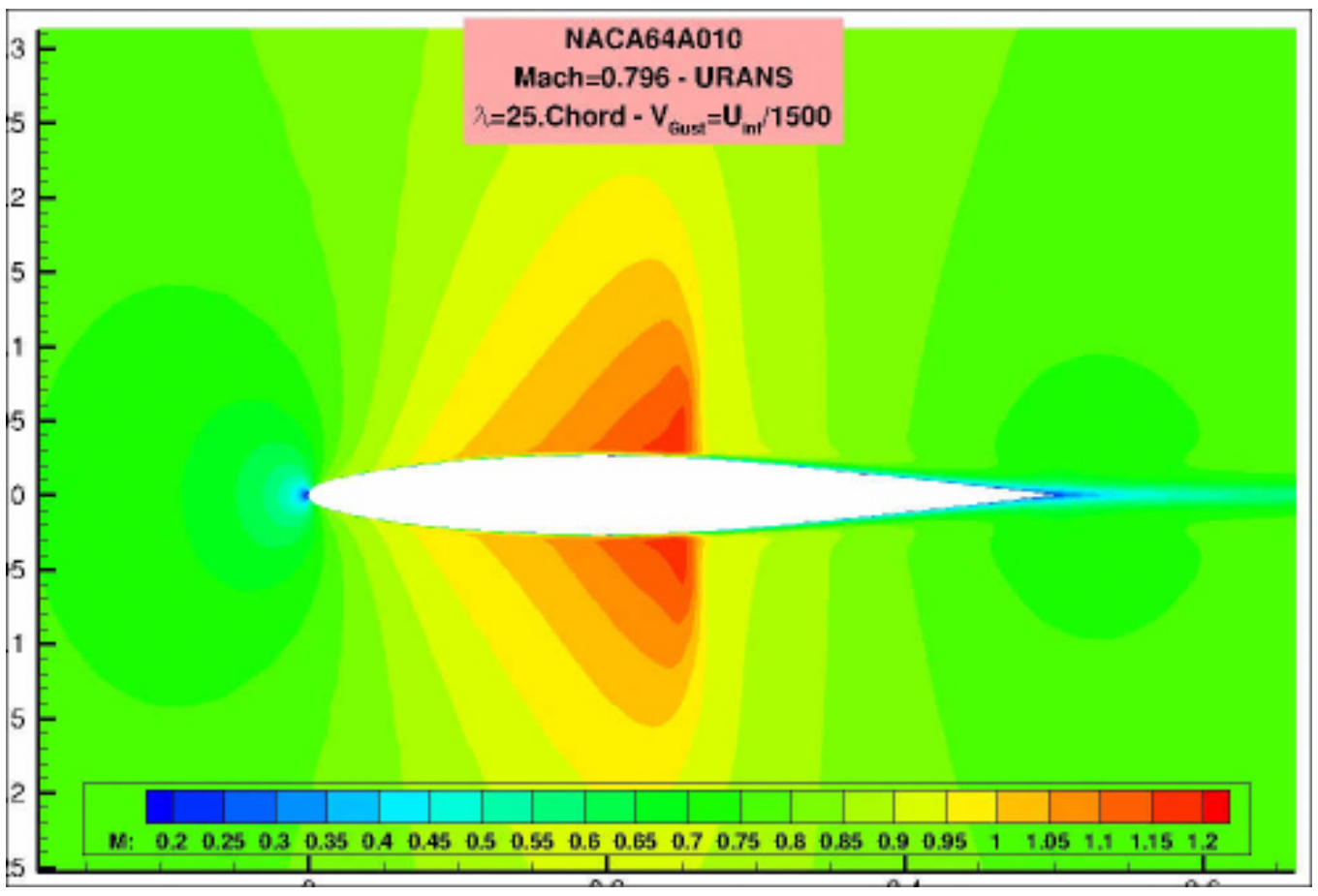

Video 2 – Mach distributions for a gust period (URANS simulation with the gust amplitude equal to U∞/1500)

## Numerical 3D benchmark for the linearized approach

This second test case is aimed at checking the validity of the linearized solver for 3D geometries in high subsonic inviscid flows. It concerns the M6 wing, for which a generic structural finite element model representing a standard spars/ribs/stiffener architecture has been built. This structural model has been used only to determine the steady state used to initialize unsteady simulations. This steady state results from a static fluid-structure coupling simulation carried out for the aerodynamic conditions defined below (Figure 18).

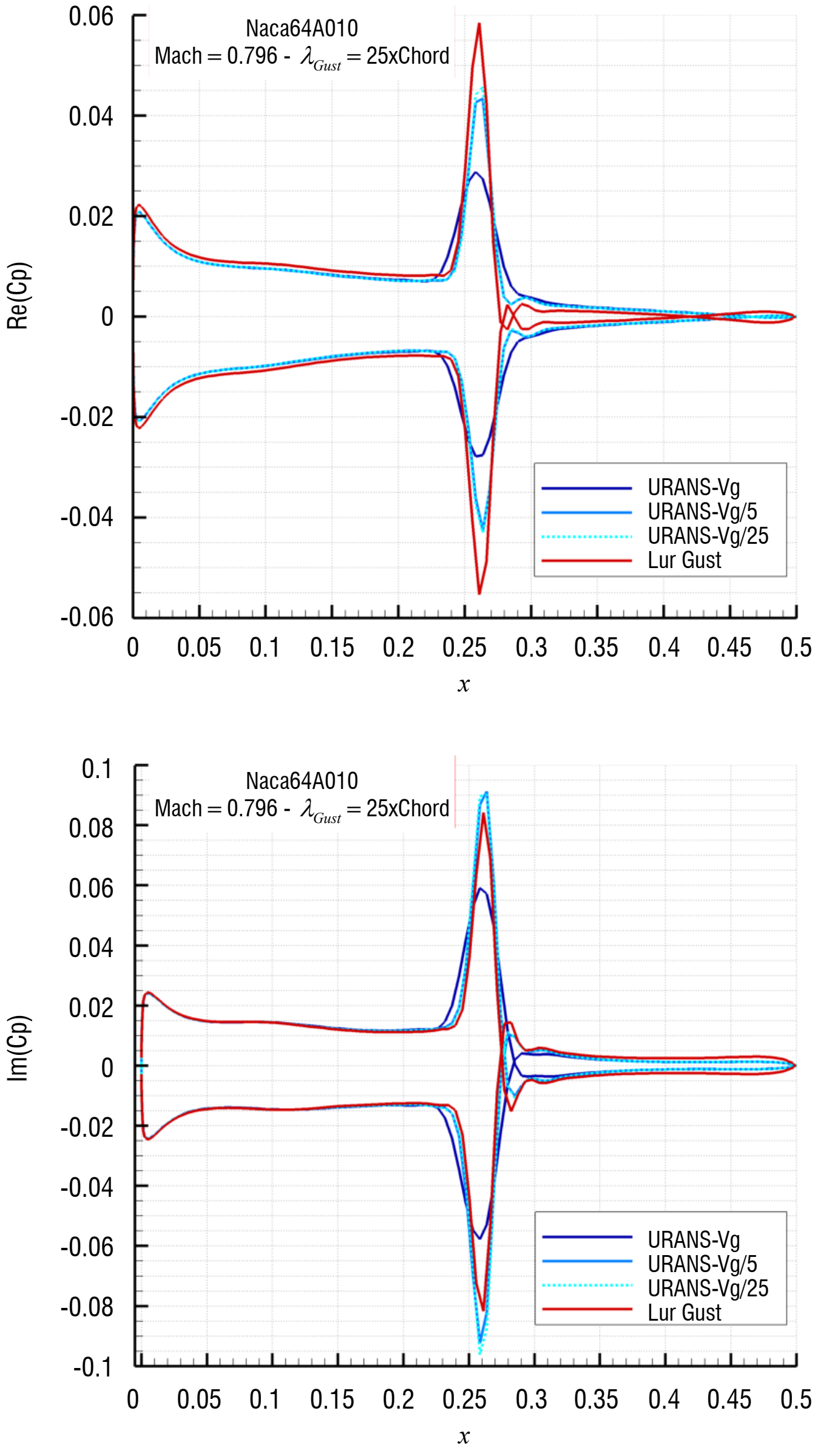

Figure 17 – Unsteady Cp distributions obtained with both the linearized solver (Lur) and the non-linear solver (URANS) (real part in the top figure, imaginary part in the bottom figure)

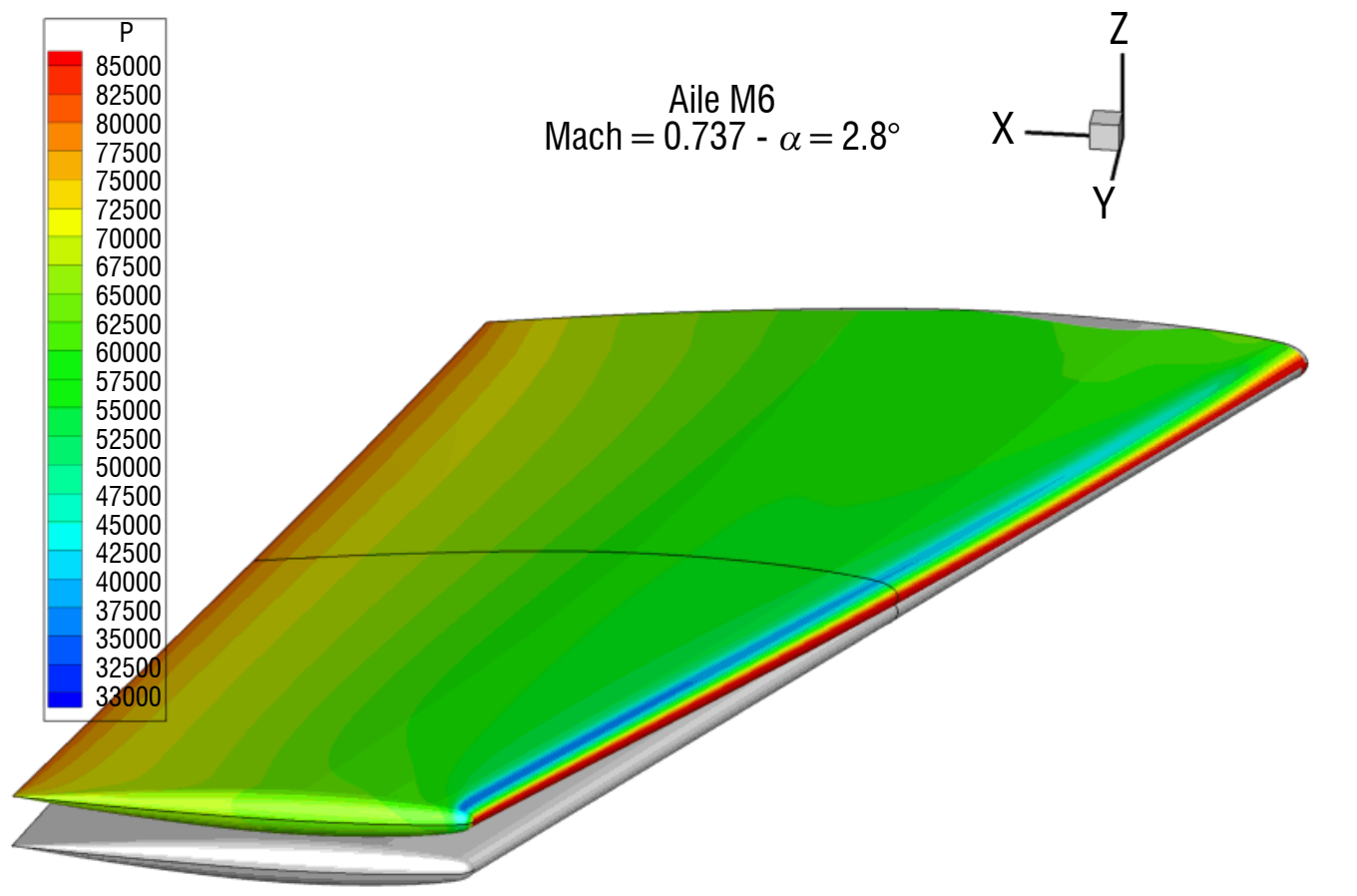

Figure 18 – Pressure distribution and wing deformations resulting from a static aeroelastic simulation

Unsteady simulations using both the non-linear and linearized solvers are performed to obtain the response to the harmonic gust defined in Table 1. Similar unsteady complex pressure (first harmonic) distributions were obtained, as can be seen in Figure 19 and Figure 20 showing the real and imaginary parts of Cp on the upper and lower surfaces. Figure 21 presents the Cp distributions on 2 span-wise sections obtained with the 2 solvers. The discrepancies, which are greater in the imaginary parts, can be explained by the amplitude of the applied gust in the non-linear simulation, which is an amplitude that is probably too high to remain in the domain of the linear unsteady perturbations.

$$\begin{cases} Mach & = & 0.734 \\ \alpha & = & 2.8^\circ \\ P_i & = & 101325\,Pa \\ T_i & = & 300\,K \\ reference\,surface & = & 0.758\,m^2 \\ mean\,chord & = & 0.805\,m \end{cases}$$

$$\begin{cases} \text{gust wavelength} & \lambda & = & 25 Chord \\ \text{gust frequency} & f & = & 12.02\,Hz \\ \text{gust amplitude} & V_g & = & \dfrac{U_\infty}{300} \\ \text{gust incidence variation} & \alpha_g & = & 0.191^\circ \end{cases}$$

Table 1 – Flight conditions and gust characteristics applied to the M6 wing

As for the 2D case, there is a significant CPU time gain for the linearized simulations: one non-linear simulation requires about 30.000 s, whereas one linearized computation requires about 4.000 s. This CPU time gain is similar to that noticed by other authors [14] using similar numerical techniques, *i.e.*, the linear system is solved using a pseudo time method with a LUSSOR implicit formulation. Nevertheless, the numerical performances could be improved up to 2 orders of magnitude [3] when most recent resolution algorithms, such as a preconditioned flexible GMRES with deflated restarting [16], are used.

## Application

Given that a gust is one of the most severe loads for an aircraft, an important issue is the gust load alleviation. The use of control laws is one way to reduce the load factor on an aircraft encountering a gust.

Control laws are built with dedicated tools using different levels of modeling for the fluid and the structure. Most often a design process uses low-fidelity aerodynamic models to synthetize control laws. However, it can be useful to check their behaviors with high-fidelity tools (efficiency, robustness and stability).

High-fidelity fluid-structure coupling simulations have thus been carried out in the case of a regional aircraft using the aileron to alleviate gust loads. The control surface has been modelled on the fluid interface, in order to obtain an interaction between the moving surface and the flow disturbance due to the gust (Figure 22).

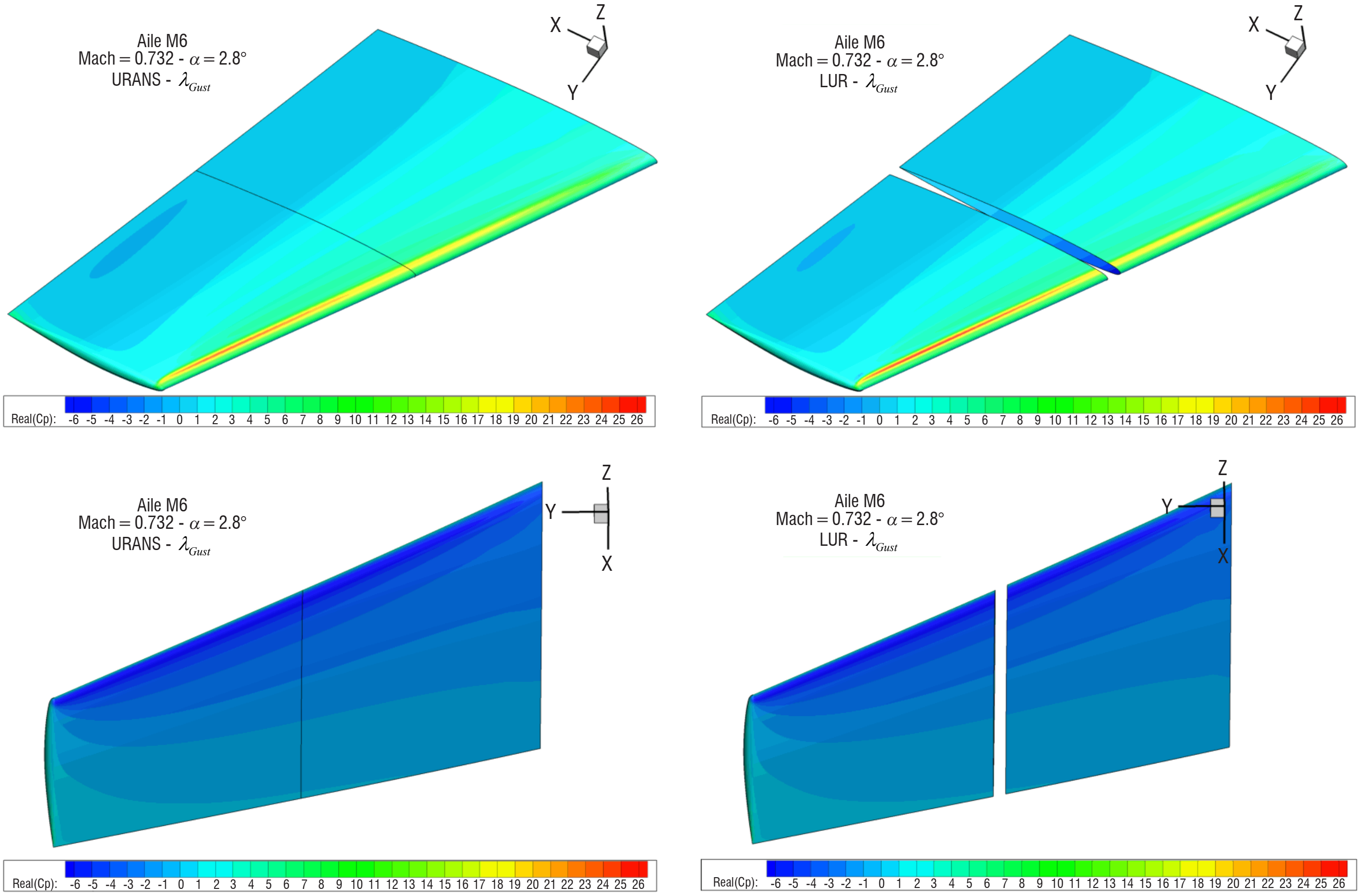

Figure 19 – M6 wing - Real part of unsteady Cp distributions resulting from a gust excitation (non-linear on the left, linearized on the right, upper surface at the top, lower surface at the bottom)

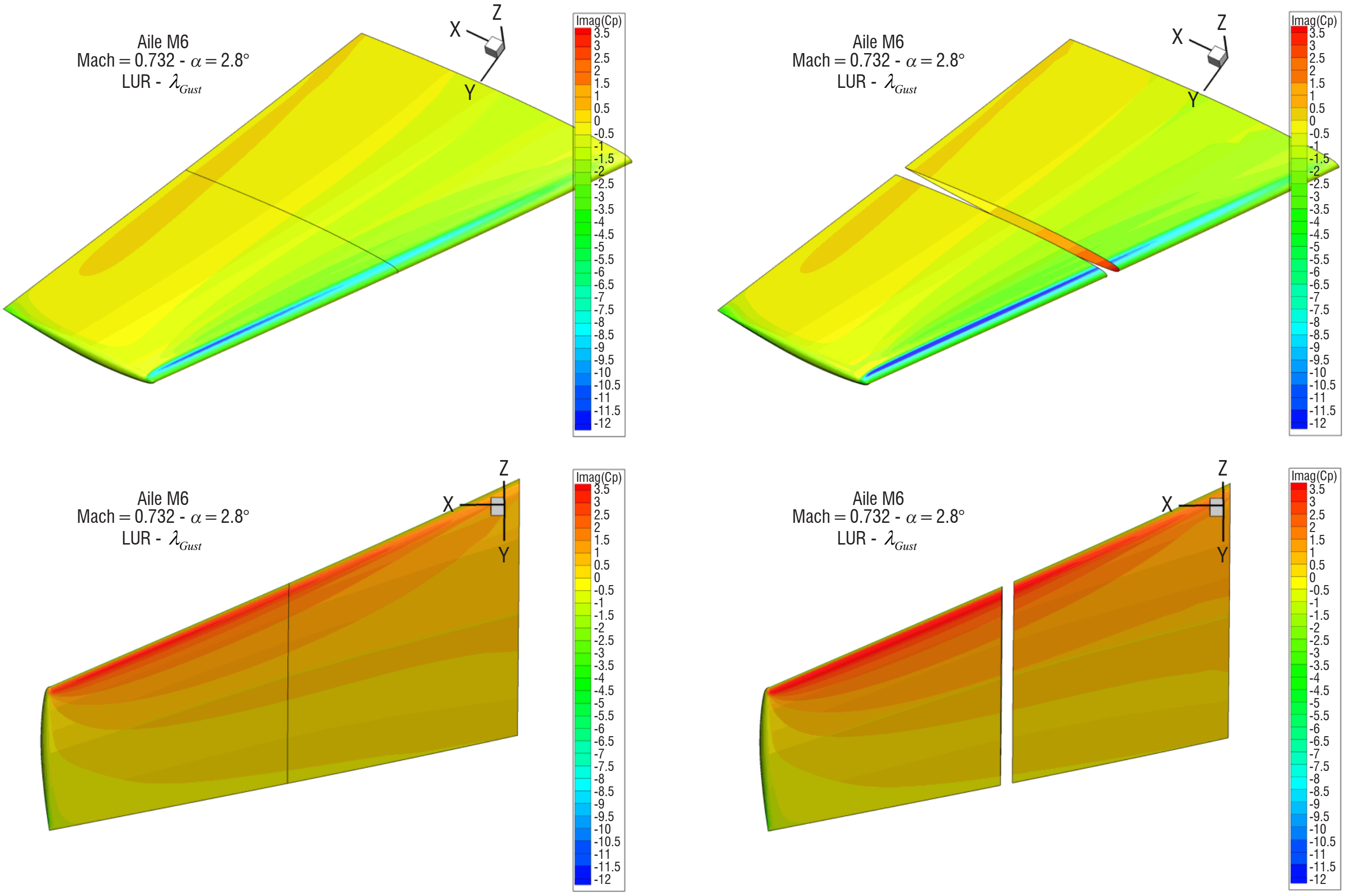

Figure 20 – M6 wing - Imaginary part of unsteady Cp distributions resulting from a gust excitation (non-linear on the left, linearized on the right, upper surface at the top, lower surface at the bottom)

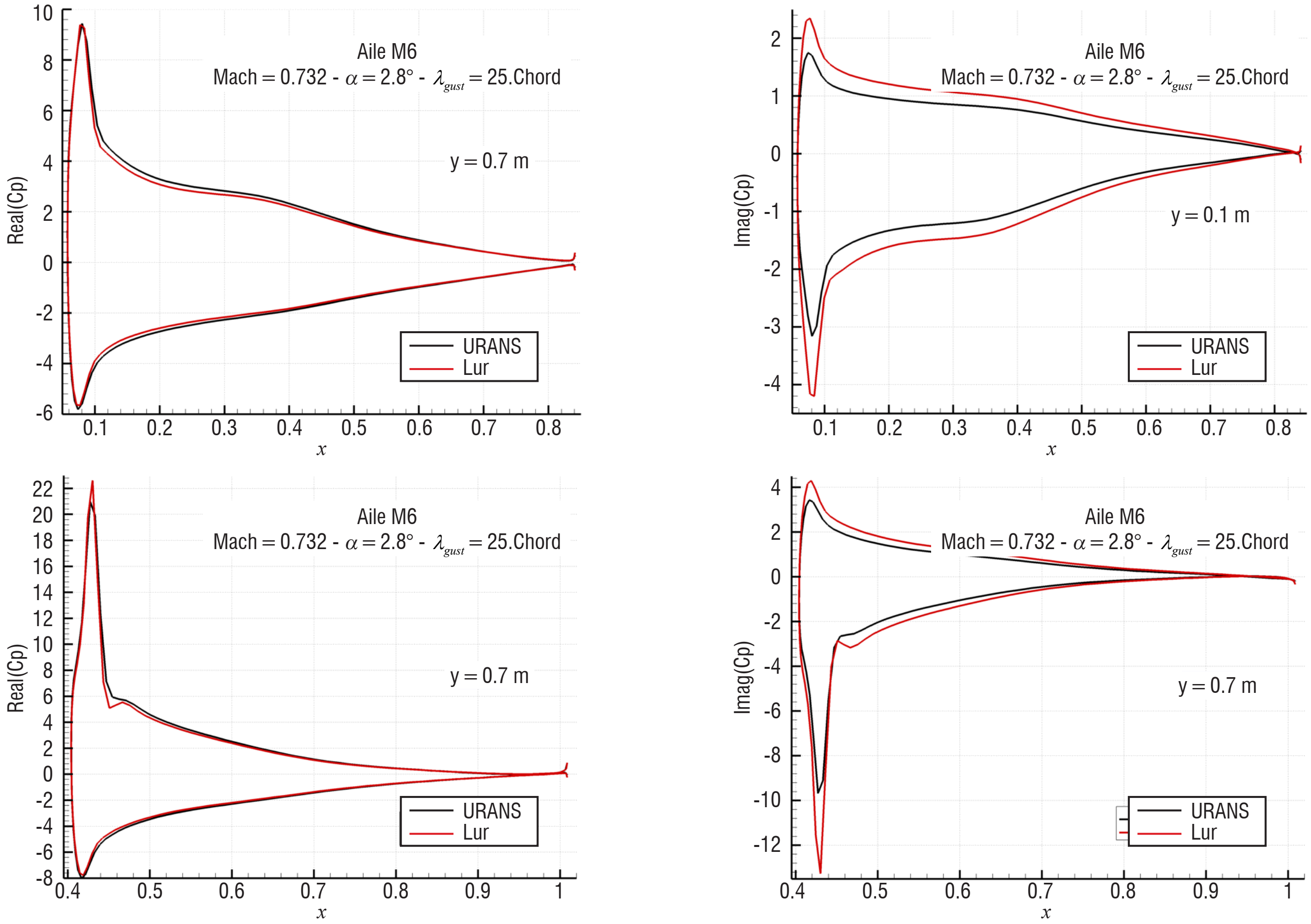

Figure 21 – Unsteady Cp distributions on 2 wing sections computed with the non-linear (URANS) and linearized (Lur) solvers

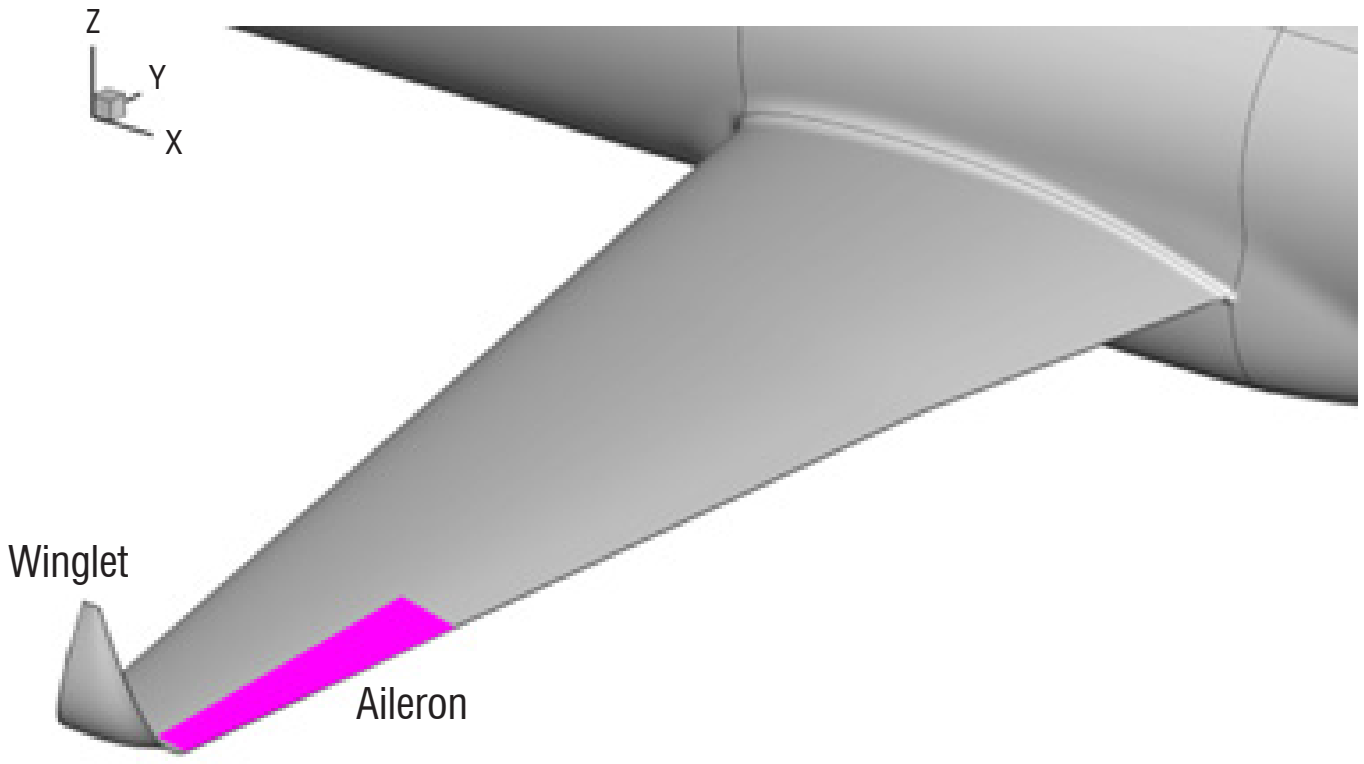

Figure 22 – Fluid interface with control surface

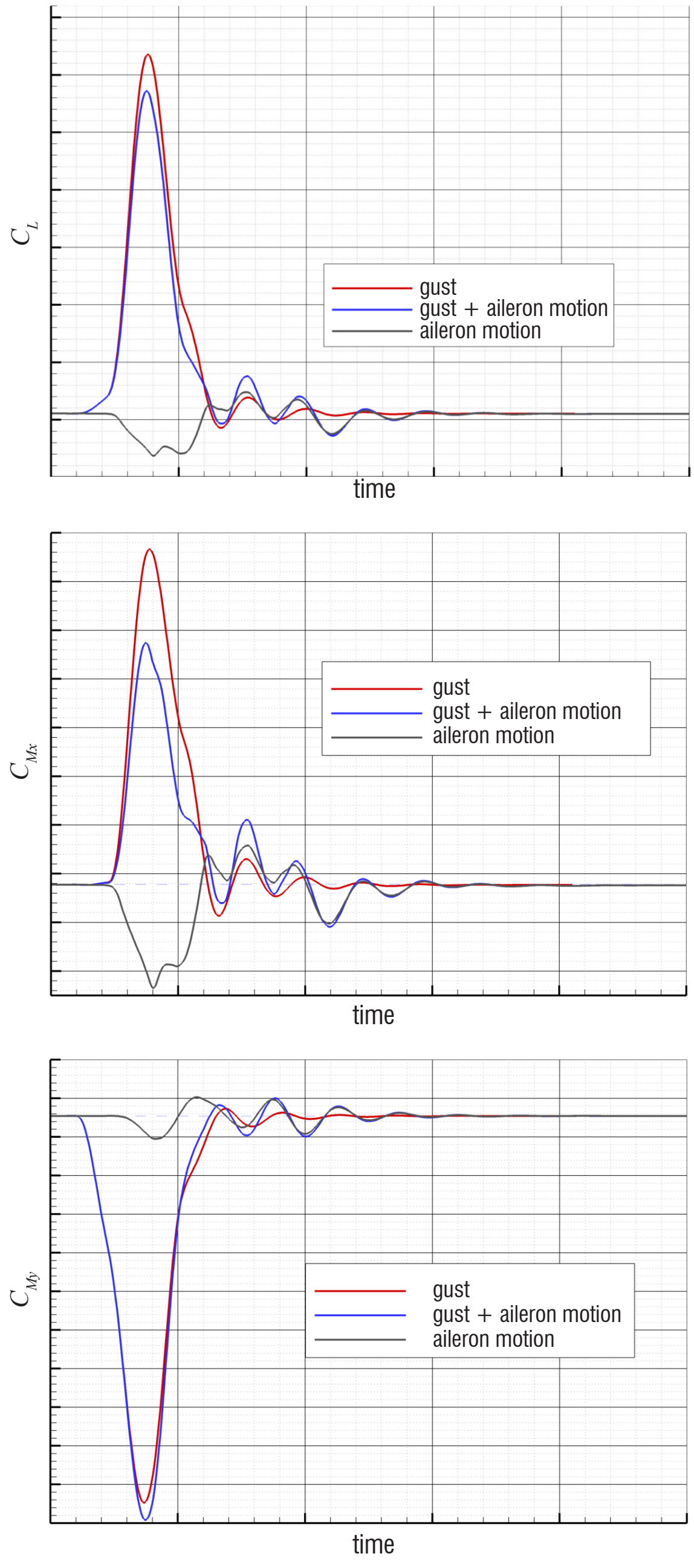

Figure 23 – Lift, rolling and pitching moment coefficient for tuned gust

A control law has been synthetized with low-fidelity tools to counteract a (1-cos) gust whose frequency is close to the first structural Eigen-frequency (first bending mode).

The gusts and corresponding control surface motions have been prescribed in a high-fidelity simulation modeling a fuselage and a wing.

Figure 23 shows the resulting time evolutions of the lift, rolling and pitching moment coefficients. An alleviation of the peak of 10.2 % due to the aileron deflection movement can then be noticed on lift, and 28 % on the rolling moment. However, a slight increase (-4.5 %) of the pitching moment is observed.

Figure 24 represents the time evolutions of the 1st (first bending mode) and 4th (first torsion mode) generalized coordinates.

The application of the aileron deflection law induces a great decrease of the main peak of the first generalized coordinate (55.8 %). The amplitude of the main peak then becomes of the same order as the amplitude of the post gust oscillations, which are strongly damped as soon as the aileron stops its deflection motion. Like this generalized coordinate, the vertical displacement of the leading edge of a wing section close to the wingtip (between the aileron and the winglet root) is 58 % alleviated by the action of the aileron motion. Indeed, this maximal displacement is equal to 0.51 m with no aileron deflection, and equal to 0.22 m with aileron motion.

The torsion modal coordinate has a time behavior similar to the generalized coordinate of the imposed aileron motion. However, the peak amplitude is almost twice that resulting from the simulation, with only the gust and without any aileron motion (-56.8 %). This 4th generalized coordinate seems to be highly sensitive to the aileron motion, and the first upward deflection of the aileron tends to amplify the peak due to the gust passage. Nevertheless, in order to obtain information from a more physical quantity about the wing deformation, the twist deformation time evolutions of the previously mentioned wing section have been extracted from the simulations. They are very different from the 4th generalized coordinates, as can be observed in Figure 25 (positive values meaning an increase in the apparent incidence of the section).

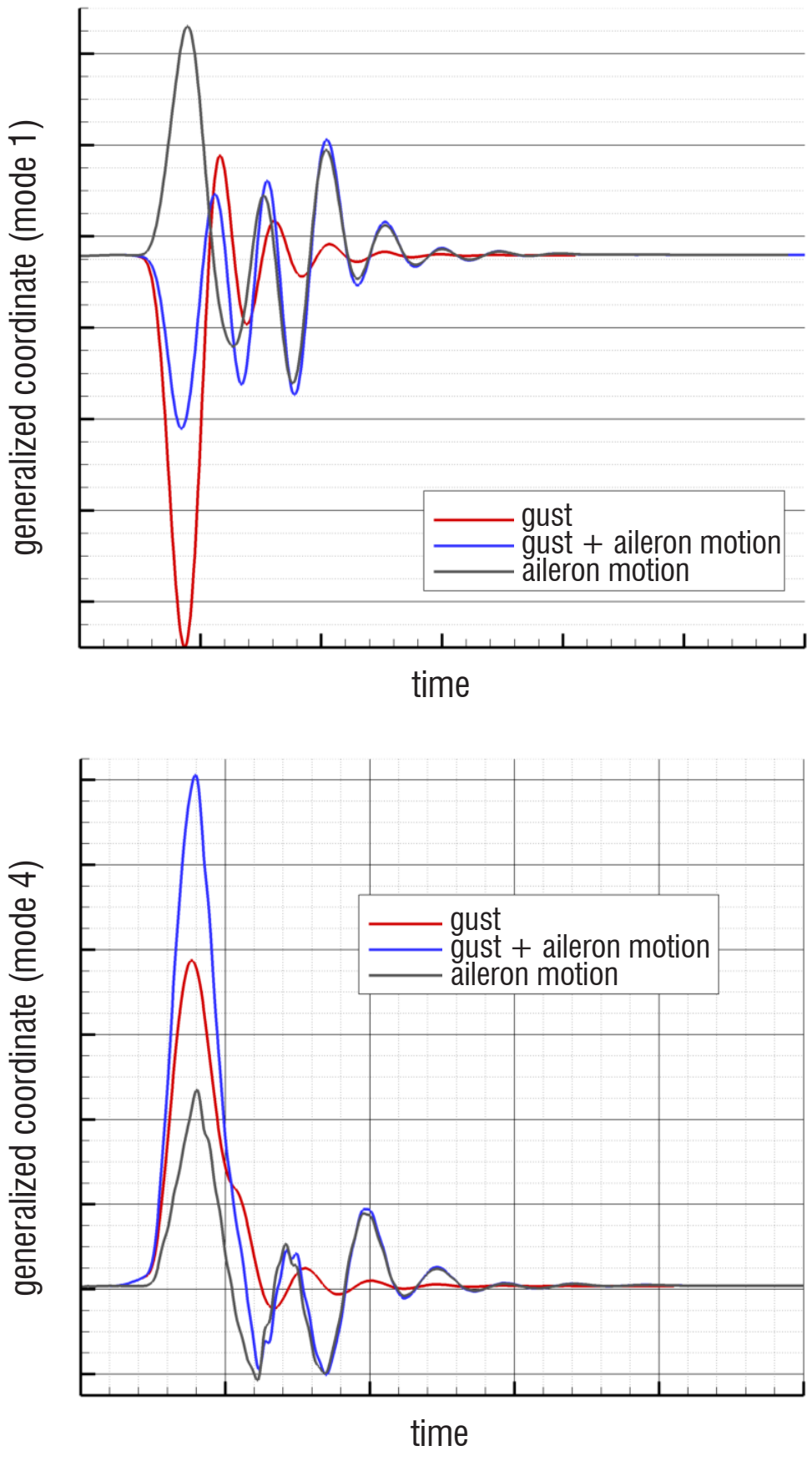

Figure 24 – Time evolutions of the 1st (1st bending) and 4th (1st torsion) modal coordinates (tuned gust)

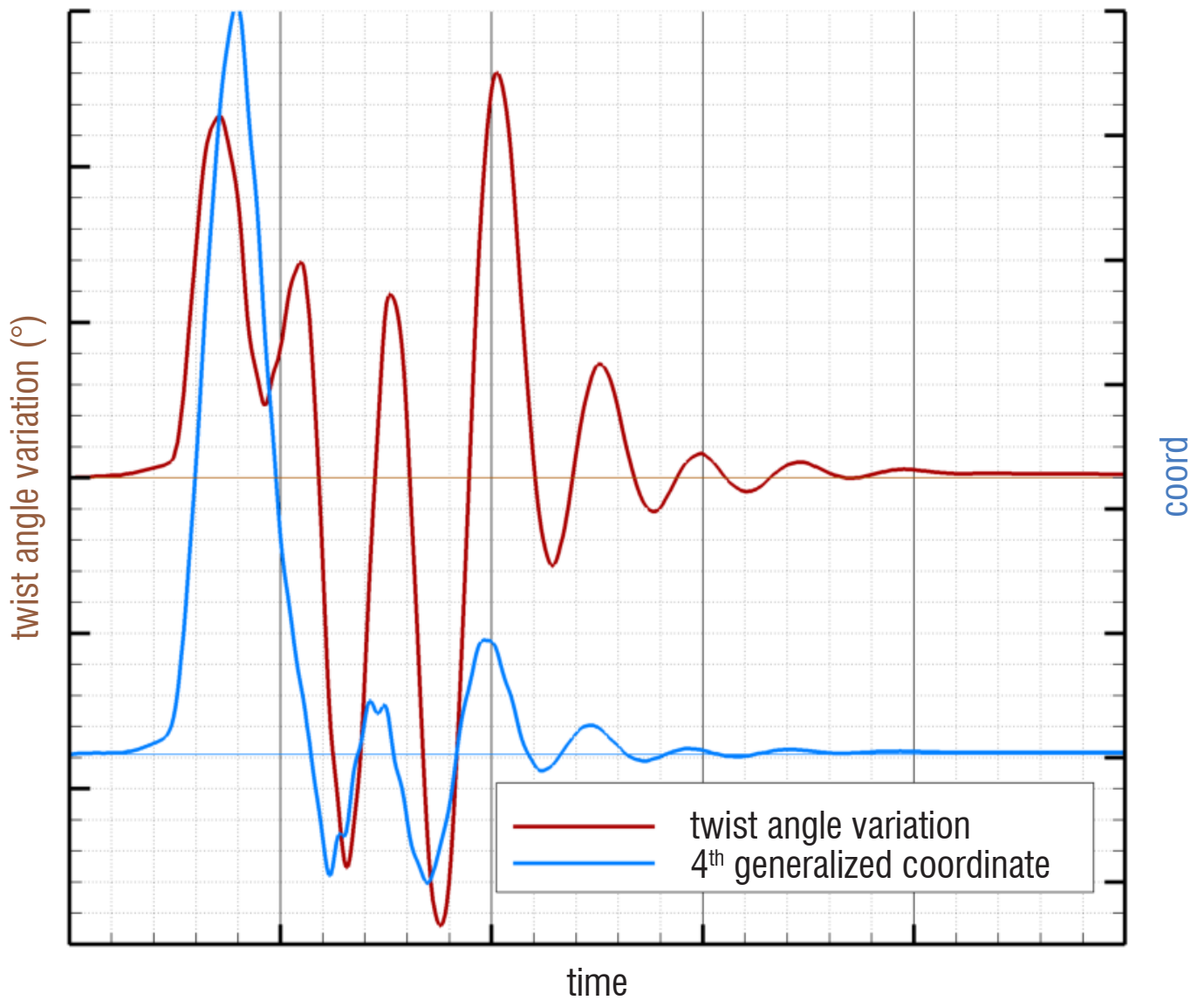

Figure 25 – Time evolutions of the twist angle variation for a wingtip section and of the 4th generalized coordinate in the case of a gust with aileron motion

This shows that the wing twist is highly influenced by modes other than the first torsion mode. Furthermore, a significant alleviation of this twist deformation due to the aileron motion is noticeable (Figure 26).

Post-gust oscillations are indeed in this case of greater amplitude than the first peak. Figure 27 shows the pressure distributions on the wing and its deformation at two different instants at which extreme wing deformations occur. The first snapshot at $t = 0.42$ s matches the instant just after the gust peak, at which the maximal value of the first generalized coordinate is reached. The second snapshot ($t = 1.0218$ s) corresponds to the lowest value of the latter generalized coordinates and to the highest downward deflection angle of the aileron after the gust passage. Finally, integrated loads have been computed with respect to time for both gust responses (with and without aileron motion) and for the aileron motion response (without gust) (Figure 28). Similar time behavior can be noticed for both the shear force and bending moment. The aileron motion induces a significant alleviation of the peak due to the gust passage (11.9 % for the shear force and 29.2 % for the bending moment). From the torsion moment point of view, the aileron motion induces the peak removal, and resulting secondary oscillations are quickly damped as soon as the aileron motion stops (these oscillations vanish after 2 s).

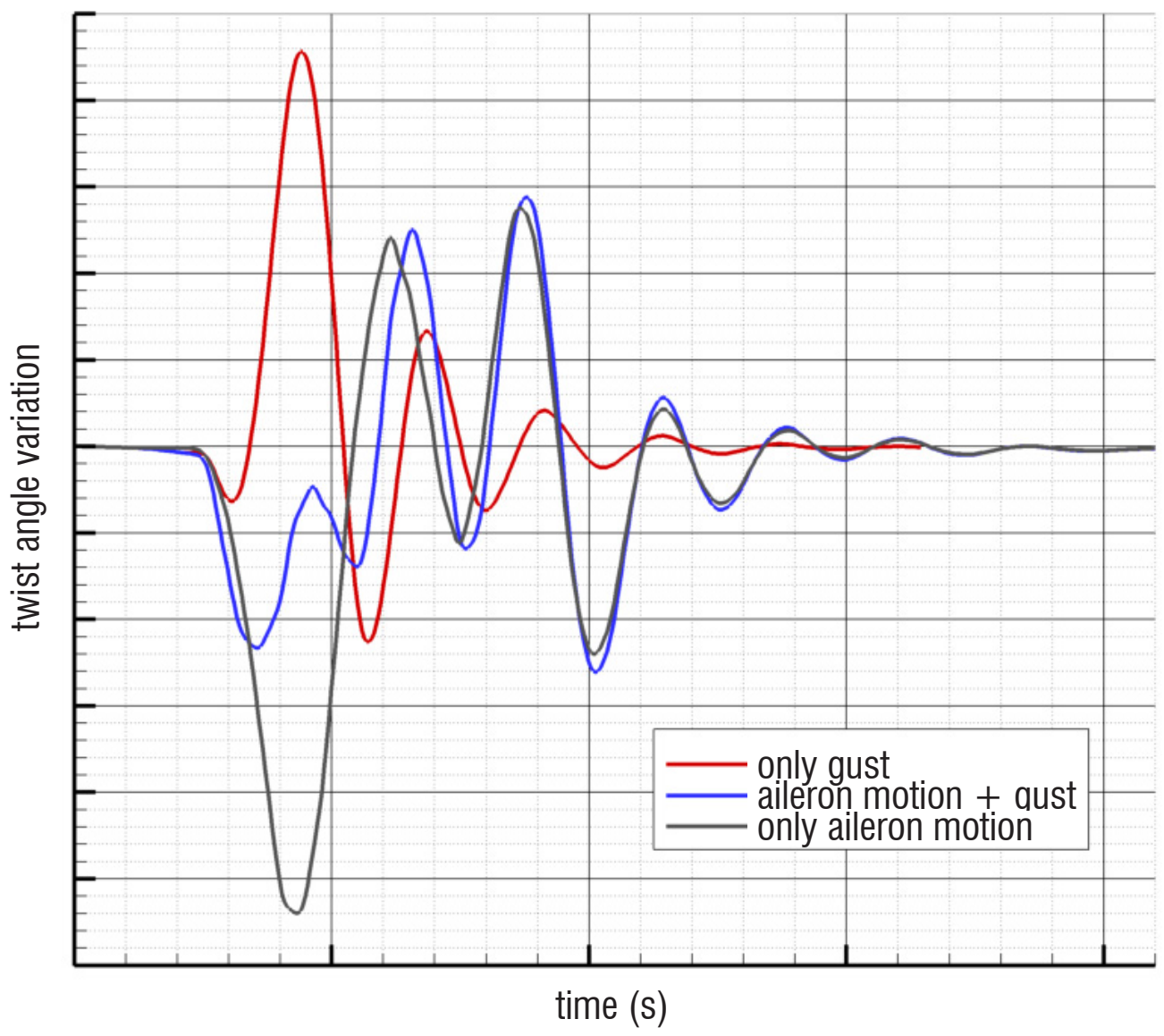

Figure 26 – Time evolutions of the twist angle variation for a section close to the wingtip

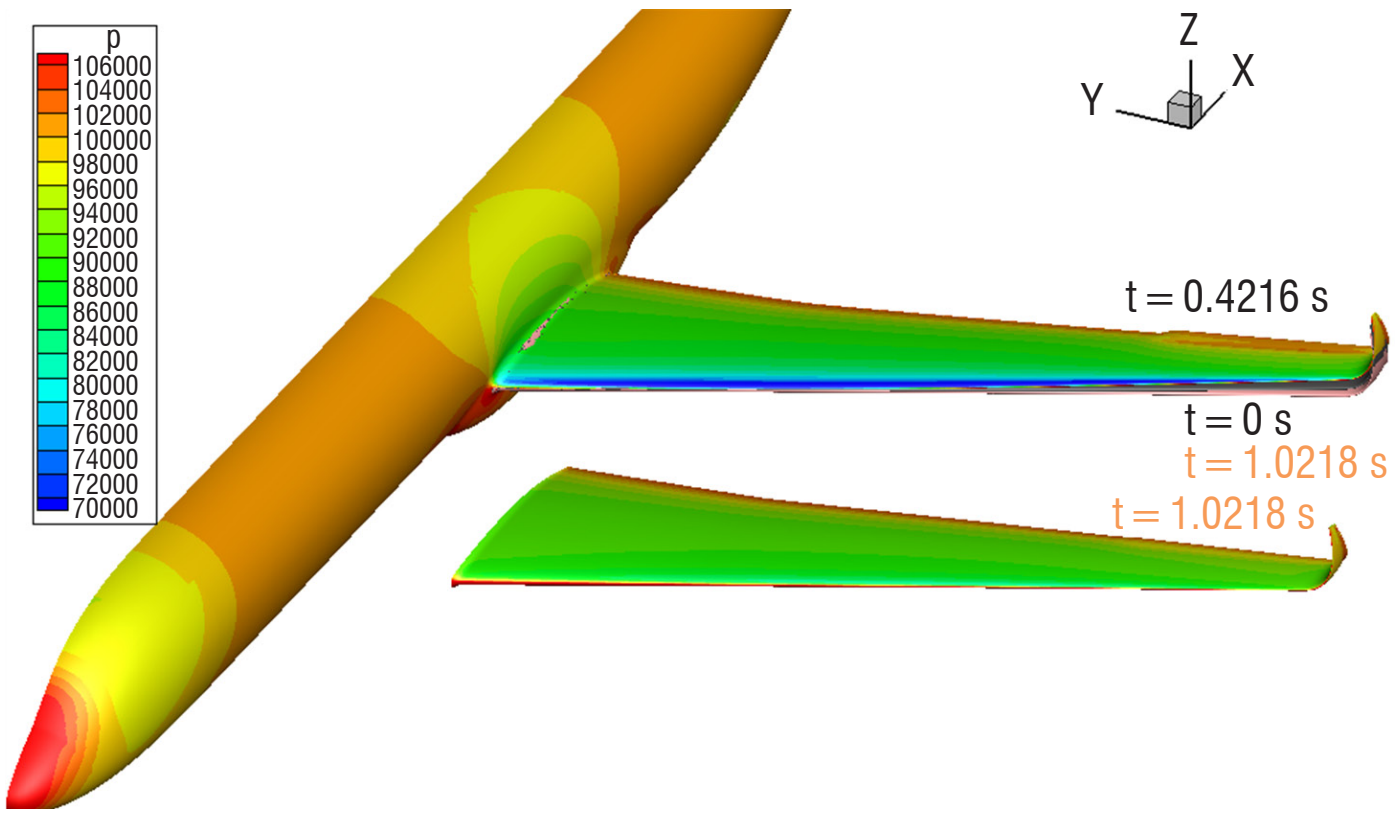

Figure 27 – Pressure distribution and wing deformation at two instants (highest upward deformation at t = 0.42 s and highest downward deformation at t = 1.02 s)

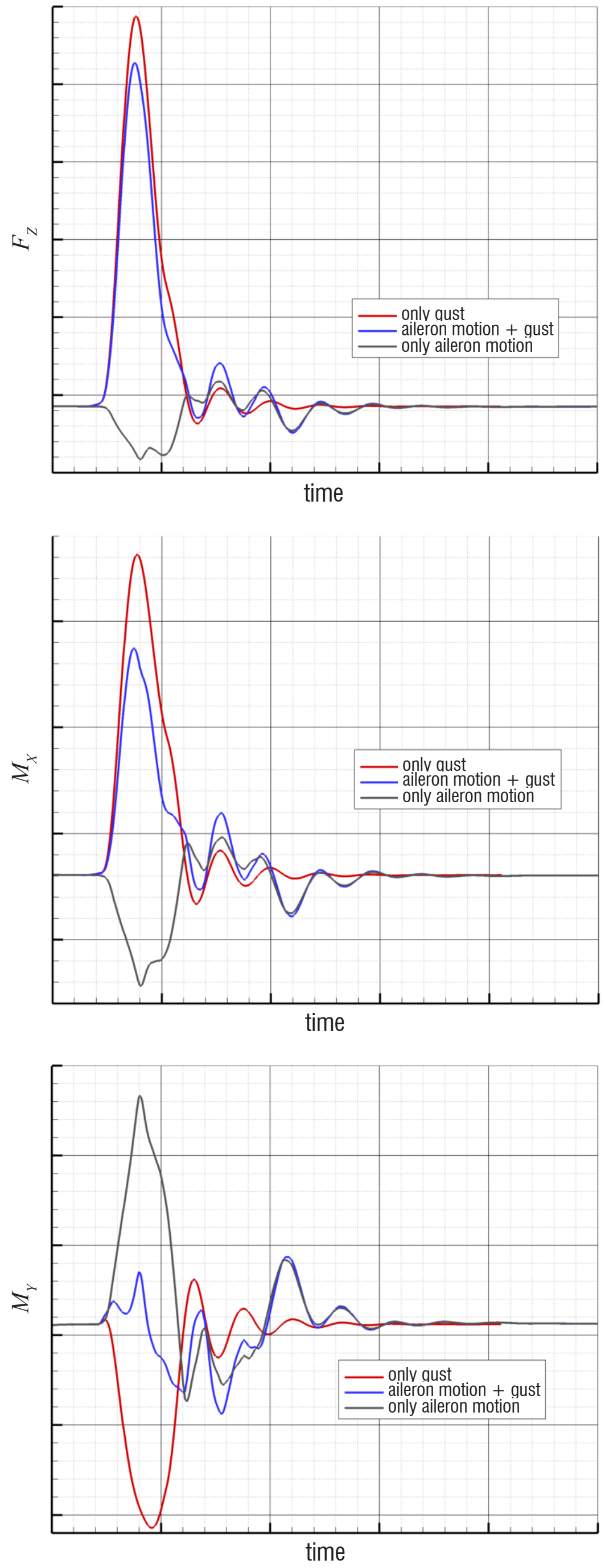

Figure 28 – Time evolution of the shear force, bending and torsion moments at the wing root

## Conclusion and Perspectives

Gusts encountered by airplanes induce loads that can be critical for some severe flight conditions, and therefore must be considered in the sizing in a structure design process. Furthermore, in the context of aircraft drag optimization and weight saving, airplane structures become increasingly flexible (large span, high wing aspect ratios). There is then a need to increase fidelity modelling to accurately assess gust loads.

High-fidelity fluid-structure coupling methodologies and simulation tools have therefore been developed to compute the response of an aircraft to a discrete gust. The first consists in modelling the flow using the URANS formulation and in solving both the structure and fluid equations in a time-consistent coupling process. The gust has therefore been modelled as an added fluid velocity field according to the Sitaraman approach. Control surface motions according to prescribed laws have also been implemented, in order to assess load alleviation and law efficiency. Such a simulation approach has first been validated by comparisons with dynamic wind tunnel experiments. A specific gust generator was designed and implemented in the wind tunnel; this generator is able to provide different kinds of time function gusts. Gust load alleviation capacities were also assessed in the case of a wing-fuselage configuration equipped with an aileron for load control.

Nevertheless, since such fluid-structure coupling simulations are very time consuming, an alternative method has been developed to obtain the aircraft response to a harmonic gust. It is based on the linearization of the URANS equations in the frequency domain.

For perspective, the free flight effects have to be accounted for in gust response simulations. Current work deals with the coupling of *elsA-Ael* to a flight dynamics model. For gust alleviation, work is underway to couple this kind of simulation to a feedback function. An update of the control law parameter based on the flow history is performed at each time step of the computation ■

**Acknowledgement**

The research leading to these results has received funding from the European Union's Seventh Framework Program (FP7/2007-2013) for the Clean Sky Joint Technology Initiative, under grant agreements CSJU-GAM-SFWA-2008-001 and CSJU-GAM-GRA-2008-001.

The studies presented in this article have been (partially) funded by Airbus, Safran, and ONERA, which are co-owners of the software *elsA*.

The authors would also like to thank AIRBUS for having provided the XRF1 aerodynamic and structural models.

The authors would like to thank Leonardo Aircraft Division for having provided data for the JTI-GRA project.

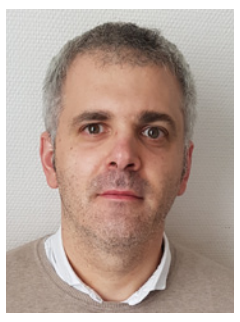

**Fabien Huvelin** graduated from the National Engineering Institute for Electronics, Computing, Telecommunications, Mathematics and Mechanics  ENSEIRB-MATMECA of Bordeaux in 2003 and obtained his PhD for his work on the fluid-structure interaction modeling by high-fidelity tool coupling in 2008. He joined the aeroelasticity department of ONERA in 2012. His main research activities are focused on the development and validation of aeroelastic modeling with high-fidelity tools.

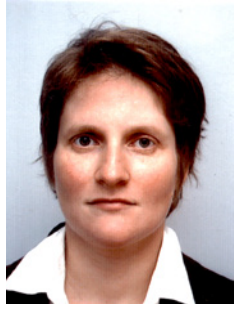

**Sylvie Dequand** graduated from ISEN Lille and specialized in "Fundamental Acoustics" at the University of Le Mans in 1997. She then obtained a PhD degree in Fluid Mechanics from the Technological University of Eindhoven in 2001. She first worked as a research assistant in Aeroacoustics at the LMS in Leuven, and then in the Automotive Department of Loughborough University. Since 2004, she has been working at ONERA as an Aeroelasticity engineer. She was first involved, together with the experimental team, in ground vibration and wind tunnel tests. Since 2008, she has been working in the Aeroelastic Modeling and Simulation unit of the Aerodynamics, Aeroelasticity and Acoustics Department of ONERA, where she is mainly in charge of fluid-structure coupling simulations and aeroelastic stability studies for civil aircraft cooperation programs.

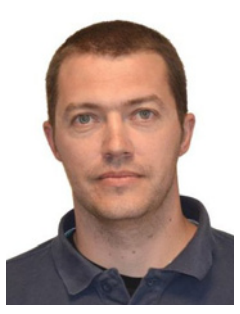

**Arnaud Lepage** graduated from the École Nationale Supérieure de Mécanique et des Microtechniques (ENSMM) of Besançon in 1998 and received a PhD degree from the University of Franche-Comté in 2002 for a thesis on experimental modal analysis in structural dynamics. Then, he joined the aeroelasticity department of ONERA as a research engineer, where he initially worked on active vibration control. Since 2006, his fields of interest have been the experimental investigation and control of fixed-wing aeroelasticity and unsteady aerodynamics (gust, buffet, and flutter).

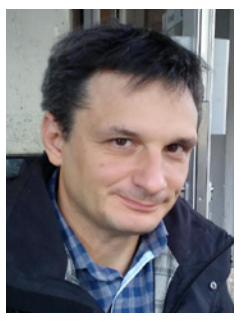

**Cédric Liauzun** graduated from the École Nationale Supérieure de Mécanique et d'Aérotechniques (ENSMA) in 1996, and has about 20 years of experience as a member of the Numerical Aeroelasticity team at ONERA, where his activity is mainly devoted to the development of numerical simulation methods for aeroelasticity and fluid-structure coupling.